\documentclass{cpbtex}
\usepackage{indentfirst}  %% ½Ú±êÌâºóµÚÒ»ÐжÎËõ½ø
\usepackage{bm}    %%%%%  ÅÅÓ¡Êýѧ·ûºÅºÚбÌåʱµ÷Óô˺ê°ü
\usepackage{graphicx}  %%%% ÎÄÖвåÈë eps ͼÐÎʱµ÷Óô˺ê°ü£¬²åͼ·½·¨ 1
\usepackage[numbers,sort&compress]{natbib}
\usepackage{color}

\usepackage{epsfig}
\usepackage{dcolumn}
\usepackage{amsmath}
\usepackage{mathrsfs}
\newtheorem{theorem}{Theorem}
\newtheorem{lemma}{Lemma}
\newtheorem{remark}{Remark}
\begin{document}

\title{Epidemic spread on one-way circular-coupled networks}

% Title should be concise; avoid abbreviations if possible; and not begin with `A', `An', `The', or `Study on'.

%\author{Zhongpu Xu$^{1}$, \ Stephen Gourley$^{2}$, \ Xinchu Fu$^{1,}$\thanks{Corresponding author. E-mail:~xcfu@shu.edu.cn}\\
%$^{1}${\small Department of Mathematics, Shanghai University, Shanghai 200444, China}\\
%$^{2}${\small Department of Mathematics, University of Surrey, Guildford, Surrey, GU2 7XH, UK}}

\author{Zhongpu Xu \ and \  Xinchu Fu\thanks{Corresponding author. E-mail:~xcfu@shu.edu.cn}\\
{\small Department of Mathematics, Shanghai University, Shanghai 200444, China}}

% The line break was forced via \\
%$^{2}${Second affiliation}  % The line break was forced via \\\Longrightarrow
%$^{3}${Third affiliation}}   % The line break was forced via \\

% 1. For Chinese authors, the name in Chinese characters should also be given. For example, Gang Liu(Áõ¸Õ), Xiao-Ming Li(ÀîÏþÃ÷)
% 2. Please ensure that every author approves the submission of the manuscript
% 3. Abbreviations should not be used in the affiliations

\date{(\today)}

\maketitle

\begin{abstract}
\noindent Real epidemic spreading networks often composed of several kinds of networks interconnected with each other, and the interrelated networks have the different topologies and epidemic dynamics. Moreover, most human diseases are derived from animals, and the zoonotic infections always spread on interconnected networks. In this paper, we consider the epidemic spreading on one-way circular-coupled network consist of three interconnected subnetworks. Here, two one-way three-layer circular interactive networks are established by introducing the heterogeneous mean-field approach method, then we get the basic reproduction numbers and prove the global stability of the disease-free equilibrium and endemic equilibrium of the models. Through mathematical analysis and numerical simulations, it is found that the basic reproduction numbers $R_0$ of the two models are dependent on the infection rates, infection periods, average degrees and degree ratios. In the first model, the network structures of the inner contact patterns have a bigger impact on $R_0$ than that of the cross contact patterns. Under the same contact pattern, the internal infection rates have greater influence on $R_0$ than the cross-infection rates. In the second model, the disease prevails in a heterogeneous network has a greater impact on $R_0$ than the disease from a homogeneous network, and the infections among the three subnetworks all play a important role in the propagation process. Numerical examples verify and expand these theoretical results very well.

\vspace{0.3cm}


\vspace{0.3cm}

\noindent\textbf{PACS:}

\end{abstract}

\section{Introduction}

Infectious diseases have been serious threats to human's health and life, and they also can bing disaster to national economy and the people's livelihood. Recently, outbreaks of different kinds of contagions (such as H7N9, SARS, Ebola and Zika) that spread around the world and thousands of people may be dead \cite{1,2}. At the same time, with the development of national economy and global warming, the emergence of new infectious diseases cause more damage to humans. So, the research of the epidemic model, epidemic spreading, preventive and control measures is a important problem, and has attracted great attentions. As early as year 1927, W.O. Kermack and A.G. Mckendrick given an analytical approach in mathematical modeling, which can predict epidemic dynamical and advise the effective measures to prevent and control the spread of disease \cite{3}.

With progress in mathematical epidemiology, researchers pay much more attention to study the feasible and practical of mathematical models of different infectious diseases. Many epidemic network models have been studied in past decades \cite{4,5}. Since the emerge of complex network theory, such as WS small-world network \cite{6} and BA scale-free network \cite{7}, then complex network into a rapid development period and becomes a powerful tool in modeling and quantifying the epidemiology. Through the modeling and computational methods (for instance mean-field approach, Markov chain and Monte Carlo simulations), researchers have obtained many research achievements about the epidemic dynamics on complex network \cite{8,9,10}. A population can be represented by a complex network, where the nodes represent individuals and the links represent various interactions among these individuals, and disease can spread along edges of the network. In fact, network modeling usually don't overly simplified as homogeneous model, and explicitly captures the diverse contact patterns of interactions in disease transmission \cite{11}. Many epidemic models are based on networks have been established to study the effect of network structures on mechanism of infection \cite{12,13,14}.

Most previous studies focus on single networks, but the fact that the real networks are too large and complex, in which the subnetwork interact with each other, such as social network, neural network, information network and transportation network \cite{15,16,17}. Then, real epidemic networks are not isolated but often composed of several kinds of networks interrelated with each other. Many diseases spread in different populations with various contact patterns and infection rates, so it is not enough to address the epidemic transmission on a single network. Thus, it is essential to study the infectious disease transmission on the multiplex networks and coupled networks.

Recently, researches in the field of epidemic spreading on  multiplex networks and coupled networks have achieved important achievements, which lead to a deeper study on epidemiology \cite{18,19,20,21,22,23}. For instance, Clara Granell et al. explored a coupled dynamical process of awareness and infection on top of multiplex networks \cite{18}. In Ref.~\cite{19}, the authors have discussed an epidemic spreading process take place on top of two interconnected complex networks. Seung-Woo Son et al. consider percolation on various interdependent or dependency networks, pointing out the close analogy to epidemic spreading on single network without these dependencies \cite{21}. In Ref.~\cite{23}, Zhen Wang et al. contrasted network-based approaches to homogeneous-mixing approaches, point out how their predictions differ, and describe the rich and often surprising behavior of disease-behavior dynamics on complex networks, and compare them to processes in statistical physics.

Now, many studies are based on two-layer interdependent network, which just contains two populations. It is a suitable method to describe the transmission of avian influenza. In Ref.~\cite{24}, the authors used a mathematical model through mean-field approximation approach to study the epidemic spreading on one-way-coupled networks comprised of two subnetworks, which can manifest the transmission of some zoonotic diseases. In fact, many real epidemics can spread among three populations, the contact pattern and infection rate in any population may be different. For example, Both HBV and HIV can spread among men, woman and children. Schistosomiasis is spread by contacting with the patient's skin and infected water, the source of the infection is the patients feces, humans and vertebrates are easily infected with schistosomiasis. So, the study of epidemic spreading on three populations has practical and important significance.

Lyme disease, an emerging zoonotic disease of natural nidus caused by Borrelia burgdorferi, transmitted by ticks, especially Ixodes scapularis \cite{24a,24b}. Human being and animals have been found to be infected with Borrelia by biting of ticks, with more than 30,000 cases reported annually in the United States alone. In China, the morbidity of the disease has also been rising in recent years. Previous studies have been reveal the pathogen transmission involves three ecological and epidemiological processes, and understand the factors that regulate the abundance and distribution of the Lyme pathogen is crucial for the effective control and prevention of the disease \cite{24c,24d,24e}. At one stage of the transmission of Lyme disease, the spread of disease between pathogens and susceptible vertebrate hosts may becomes a cycle. Here, we first establish the epidemic spreading on a one-way three-layer circular-coupled networks, in which three subnetworks denote three different populations, and the spread of infectious disease is one-way circular process.

As a matter of fact, the latest survey says that a majority of human pathogens are zoonotic or originated as zoonoses before affecting humans. Some diseases are caused by zoonoses, such as Black death, Rabies and Spanish influenza \cite{25,26,27}. Many infectious diseases transmit in human society through vectors, and the vector as an infective media. For example, malaria is a life-threatening disease caused by parasites that are transmitted to people through the bites of infected mosquitoes. The Black death is the modern name given to the deadly epidemic disease spread by rat fleas across Europe in the 14th century. The emergence of complex zoonotic infections pose a great threat to the public healthy, it is urge to study the modeling in prevention and control of zoonotic infections.

Recently, some researches about human-animal coupled epidemic model on complex network has drawn attention \cite{28,29,30,31}. Especially, Hongjing Shi et al. set up a new susceptible-infected-susceptible (SIS) model with infective medium, which describes epidemics transmitted by infective media on various complex networks \cite{28}. Yi Wang et al. proposed and analyzed a modified SIS model with an infective vector on complex networks, which incorporates some infectious diseases that are not only transmitted by a vector, but also spread by direct contacts between humans \cite{29}.  In Ref.~\cite{31}, the authors established an epidemic time-evolution model of some zoonotic infections by the mean-field approximation and its global dynamics are investigated.

Some previous researches are mainly about an epidemic model with infective media connecting two separated networks of populations. Here, we consider a one-way three-layer circular-coupled networks with three different vectors (infective mediums), in which between any two subnetworks has an infective medium. The study of the special epidemic model provides a new sight on the zoonotic infections on interconnected networks.

In this paper, we intend to use analytical and numerical methods to discuss the epidemic transmission on the two types of one-way three-layer coupled networks. In Section $2$, a one-way three-layer circular-coupled network is introduced by heterogeneous mean-field approach method, then find out the basic reproduction number and prove the global stability of the disease-free equilibrium and endemic equilibrium. In Section $3$, we consider a one-way three-layer circular-coupled network with infective mediums. Similarly, we also calculate the basic reproduction number, then study the global dynamics of this model. In Section $4$, we perform some numerical simulations to verify and supplement the theory of two new models. Conclusion is given in Section $5$.

\section{One-way circular-coupled network}

\begin{figure}[]
\begin{center}
\includegraphics[width=9cm,height=5cm]{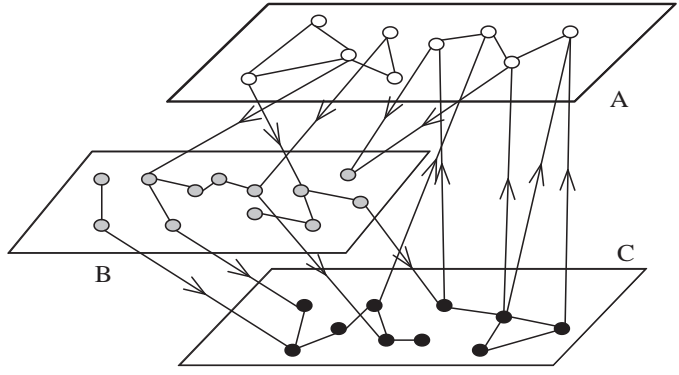}
\caption{A one-way circular-coupled network with three subnetworks.}\label{e001}
\end{center}
\end{figure}

In this section, we consider a one-way circular-coupled network system. The whole network include three subnetworks, $A$, $B$ and $C$. In each subnetwork, nodes represent individuals and the links represent various contacts among these individuals, and disease can spread along edges of the network. In the network, there are cross contacts between $A$ and $B$, $B$ and $C$, $A$ and $C$, i.e., the nodes in $A$, $B$ and $C$ all have three types of degree. Moreover, the disease spreading in three subnetworks is one-way, that means the nodes in $A$ can infect the nodes in $B$, similarly, network $B$ can infect $C$, network $C$ can infect $A$, but not the reverse. The structure of a one-way circular-coupled network is shown in Fig.\ref{e001}.

In each subnetwork, we consider the classical susceptible-infected-susceptible (SIS) epidemic model as the topological structure, so each individual can be one of two distinct states at each time: susceptible(S) and infected(I). To describe the epidemic spreading process of the whole network, we let $P_A(i,j,k)$ as the probabilities of choosing a random node in $A$ with $i$ links within $A$, $j$ links connecting $A$ and $B$, and $k$ links connecting $A$ and $C$; similarly, $P_B(i,j,k)$ as the probabilities of choosing a random node in $B$ with $i$ links connecting $B$ and $A$, $j$ links within $B$, and $k$ links connecting $B$ and $C$; $P_C(i,j,k)$ as the probabilities of choosing a random node in $C$ with $i$ links connecting $C$ and $A$, $j$ links connecting $C$ and $B$, and $k$ links within $C$. And others parameters about the network are given in Table \ref{e002}.

\begin{tabular}{ c|l }
  \hline
    Parameter & Meaning ($X=A, B$ or $C$) \\
  \hline
  $N_{i,j,k}^X$                     & Number of nodes in $X$ with $(i,j,k)$-degree\\
  $S_{i,j,k}^X$(or $I_{i,j,k}^X$)   & Number of susceptible (infected) nodes in $X$ with $(i,j,k)$-degree\\
  $P_X(i,j,k)$                      & Probability of choosing a random node in $X$ with $(i,j,k)$-degree\\
  $n_{11}$(or $n_{12}$ or $n_{13}$) & Maximum degree value of nodes in $A$ connecting $A$ (or $B$ or $C$)\\
  $n_{21}$(or $n_{22}$ or $n_{23}$) & Maximum degree value of nodes in $B$ connecting $A$ (or $B$ or $C$) \\
  $n_{31}$(or $n_{32}$ or $n_{33}$) & Maximum degree value of nodes in $C$ connecting $A$ (or $B$ or $C$) \\
  $\langle{k}\rangle_{11}$(or $\langle{k}\rangle_{12}$ or $\langle{k}\rangle_{13}$) & Average degree value of nodes in $A$ connecting $A$ (or $B$ or $C$)\\
  $\langle{k}\rangle_{21}$(or $\langle{k}\rangle_{22}$ or $\langle{k}\rangle_{23}$) & Average degree value of nodes in $B$ connecting $A$ (or $B$ or $C$) \\
  $\langle{k}\rangle_{31}$(or $\langle{k}\rangle_{32}$ or $\langle{k}\rangle_{33}$) & Average degree value of nodes in $C$ connecting $A$ (or $B$ or $C$) \\
  \hline
\end{tabular}\label{e002} \\

According to Table $2$,  we can get the total number of susceptible nodes, infected nodes and all nodes in subnetwork $A$, $B$ and $C$, respectively, which are
 $$ S^A=\sum\limits_{i=0}^{n_{11}}\sum\limits_{j=0}^{n_{12}}\sum\limits_{k=0}^{n_{13}}S_{i,j,k}^A,\ I^A=\sum\limits_{i=0}^{n_{11}}\sum\limits_{j=0}^{n_{12}}\sum\limits_{k=0}^{n_{13}}I_{i,j,k}^A, \ N^A=S^A+I^A;$$

 $$ S^B=\sum\limits_{i=0}^{n_{21}}\sum\limits_{j=0}^{n_{22}}\sum\limits_{k=0}^{n_{23}}S_{i,j,k}^B,\ I^B=\sum\limits_{i=0}^{n_{21}}\sum\limits_{j=0}^{n_{22}}\sum\limits_{k=0}^{n_{23}}I_{i,j,k}^B, \ N^B=S^B+I^B;$$

 $$ S^C=\sum\limits_{i=0}^{n_{31}}\sum\limits_{j=0}^{n_{32}}\sum\limits_{k=0}^{n_{33}}S_{i,j,k}^C,\ I^C=\sum\limits_{i=0}^{n_{31}}\sum\limits_{j=0}^{n_{32}}\sum\limits_{k=0}^{n_{33}}I_{i,j,k}^C, \ N^C=S^C+I^C.$$

 The joint degree distributions are
 $$P_A(i,j,k)=\frac{N_{i,j,k}^A}{N^A},\ P_B(i,j,k)=\frac{N_{i,j,k}^B}{N^B},\ P_C(i,j,k)=\frac{N_{i,j,k}^C}{N^C},$$
 the marginal degree distributions are
 $$P_A(i,\cdot,\cdot)=\sum\limits_{j=0}^{n_{12}}\sum\limits_{k=0}^{n_{13}}P_A(i,j,k),\ P_A(\cdot,j,\cdot)=\sum\limits_{i=0}^{n_{11}}\sum\limits_{k=0}^{n_{13}}P_A(i,j,k),\ P_A(\cdot,\cdot,k)=\sum\limits_{i=0}^{n_{11}}\sum\limits_{j=0}^{n_{12}}P_A(i,j,k);$$

 $$P_B(i,\cdot,\cdot)=\sum\limits_{j=0}^{n_{22}}\sum\limits_{k=0}^{n_{23}}P_B(i,j,k),\ P_B(\cdot,j,\cdot)=\sum\limits_{i=0}^{n_{21}}\sum\limits_{k=0}^{n_{23}}P_B(i,j,k),\ P_B(\cdot,\cdot,k)=\sum\limits_{i=0}^{n_{21}}\sum\limits_{j=0}^{n_{22}}P_B(i,j,k);$$

 $$P_C(i,\cdot,\cdot)=\sum\limits_{j=0}^{n_{32}}\sum\limits_{k=0}^{n_{33}}P_C(i,j,k),\ P_C(\cdot,j,\cdot)=\sum\limits_{i=0}^{n_{31}}\sum\limits_{k=0}^{n_{33}}P_C(i,j,k),\ P_C(\cdot,\cdot,k)=\sum\limits_{i=0}^{n_{31}}\sum\limits_{j=0}^{n_{32}}P_C(i,j,k),$$
and the average degree values($\alpha=1$) and the second moments of degree($\alpha=2$) are
$$\langle{k^\alpha}\rangle_{11}=\sum\limits_{i=0}^{n_{11}}i^\alpha P_A(i,\cdot,\cdot),\ \langle{k^\alpha}\rangle_{12}=\sum\limits_{j=0}^{n_{12}}j^\alpha P_A(\cdot,j,\cdot),\ \langle{k^\alpha}\rangle_{13}=\sum\limits_{k=0}^{n_{13}}k^\alpha P_A(\cdot,\cdot,k);$$

$$\langle{k^\alpha}\rangle_{21}=\sum\limits_{i=0}^{n_{21}}i^\alpha P_B(i,\cdot,\cdot),\ \langle{k^\alpha}\rangle_{22}=\sum\limits_{j=0}^{n_{22}}j^\alpha P_B(\cdot,j,\cdot),\ \langle{k^\alpha}\rangle_{23}=\sum\limits_{k=0}^{n_{23}}k^\alpha P_B(\cdot,\cdot,k);$$

$$\langle{k^\alpha}\rangle_{31}=\sum\limits_{i=0}^{n_{31}}i^\alpha P_C(i,\cdot,\cdot),\ \langle{k^\alpha}\rangle_{32}=\sum\limits_{j=0}^{n_{32}}j^\alpha P_C(\cdot,j,\cdot),\ \langle{k^\alpha}\rangle_{33}=\sum\limits_{k=0}^{n_{33}}k^\alpha P_C(\cdot,\cdot,k).$$

For simplicity, we don't consider the birth, death and immunization of individuals in the whole model, so the subnetwork sizes $N^A, N^B$ and $N^C$ are all constant. Hence, notice that the total number of links in $A$ connecting to $B$ equals the number of links in $B$ connecting to $A$, it is similar to the number of links in $B$ and $C$, the number of links in $C$ and $A$, i.e., $\sum\limits_{j=0}^{n_{12}}jN_{\cdot,j,\cdot}^A=\sum\limits_{i=0}^{n_{21}}iN_{i,\cdot,\cdot}^B,\ \sum\limits_{k=0}^{n_{23}}kN_{\cdot,\cdot,k}^B=\sum\limits_{j=0}^{n_{32}}jN_{\cdot,j,\cdot}^C,\  \sum\limits_{i=0}^{n_{31}}iN_{i,\cdot,\cdot}^C=\sum\limits_{k=0}^{n_{13}}kN_{\cdot,\cdot,k}^A,$ so we obtain
$$N^A\langle{k}\rangle_{12}=N^B\langle{k}\rangle_{21},\ N^B\langle{k}\rangle_{23}=N^C\langle{k}\rangle_{32},\ N^A\langle{k}\rangle_{13}=N^C\langle{k}\rangle_{31}.$$

Hence, we construct a one-way circular-coupled epidemic network model, and present SIS epidemic model on the network. An susceptible node in subnetwork $A$(or $B$ or $C$) can be infected with rate $\lambda_{11}, \lambda_{22}$ and $\lambda_{33}$ if it is connected to an infected node in subnetwork $A$(or $B$ or $C$), respectively. In the one-way epidemic network, a susceptible node in subnetwork $B$ has probability $\lambda_{12}$ of contagion with an infected node in subnetwork $A$, a susceptible node in subnetwork $C$ can be infected with probability $\lambda_{23}$ if it is connected to an infected node in subnetwork $B$, and the probability $\lambda_{31}$ of a susceptible node in subnetwork $A$ can be infected by an infected node in subnetwork $C$. Each infected node in subnetworks $A$, $B$ and $C$ becomes susceptible with rate $\mu_1, \mu_2$ and $\mu_3$, respectively.

Let $s_{e,f,g}^A=\frac{S_{e,f,g}^A}{N_{e,f,g}^A},\ \rho_{e,f,g}^A=\frac{I_{e,f,g}^A}{N_{e,f,g}^A}$ as the corresponding densities of the subnetwork $A$, similar to the definition of $s_{l,m,n}^B, \rho_{l,m,n}^B, s_{x,y,z}^C$ and $\rho_{x,y,z}^C$. Without loss of generality, we suppose the relationships between infected and susceptible nodes are uncorrelated, which means the connectivity of any node in network is irrelevant to the connectivity of its neighbors.

Now, we give a dynamical mean-field reaction rate equations of the epidemic network, which is composed of $[(n_{11}+1)(n_{12}+1)(n_{13}+1)+(n_{21}+1)(n_{22}+1)(n_{23}+1)+(n_{31}+1)(n_{32}+1)(n_{33}+1)]$-dimensional ordinary differential equations:

\begin{equation}\label{e01}
\left\{\begin{array}{ll}
\frac{d\rho_{e,f,g}^A(t)}{dt}=\lambda_{11}e(1-\rho_{e,f,g}^A(t))\Theta_{11}(t)+\lambda_{31}g(1-\rho_{e,f,g}^A(t))\Theta_{31}(t)-\mu_1\rho_{e,f,g}^A(t),  \\
\frac{d\rho_{l,m,n}^B(t)}{dt}=\lambda_{12}l(1-\rho_{l,m,n}^B(t))\Theta_{12}(t)+\lambda_{22}m(1-\rho_{l,m,n}^B(t))\Theta_{22}(t)-\mu_2\rho_{l,m,n}^B(t), \\
\frac{d\rho_{x,y,z}^C(t)}{dt}=\lambda_{23}y(1-\rho_{x,y,z}^C(t))\Theta_{23}(t)+\lambda_{33}z(1-\rho_{x,y,z}^C(t))\Theta_{33}(t)-\mu_3\rho_{x,y,z}^C(t),
\end{array}\right.
\end{equation}
where $e=0,1,\cdots,n_{11},\ f=0,1,\cdots,n_{12},\ g=0,1,\cdots,n_{13},\ l=0,1,\cdots,n_{21},\ m=0,1,\cdots,n_{22},\ n=0,1,\cdots,n_{23},\ x=0,1,\cdots,n_{31},\ y=0,1,\cdots,n_{32},\ z=0,1,\cdots,n_{33},$  and the meanings of $\Theta_{ij}$ for the situation $(i,j)=(1,1),(1,2),(2,2),(2,3),(3,1)$ and $(3,3)$ are as follows:
$$\Theta_{11}(t)=\frac{\sum\limits_{i=0}^{n_{11}}i\bigg(\sum\limits_{j=0}^{n_{12}}\sum\limits_{k=0}^{n_{13}}I_{i,j,k}^A(t)\bigg)}
{\sum\limits_{i=0}^{n_{11}}i\bigg(\sum\limits_{j=0}^{n_{12}}\sum\limits_{k=0}^{n_{13}}N_{i,j,k}^A\bigg)}=\frac{1}{\langle{k}\rangle_{11}}
\sum\limits_{i=0}^{n_{11}}\sum\limits_{j=0}^{n_{12}}\sum\limits_{k=0}^{n_{13}}iP_A(i,j,k)\rho_{i,j,k}^A(t),$$

$$\Theta_{12}(t)=\frac{1}{\langle{k}\rangle_{12}}
\sum\limits_{i=0}^{n_{11}}\sum\limits_{j=0}^{n_{12}}\sum\limits_{k=0}^{n_{13}}jP_A(i,j,k)\rho_{i,j,k}^A(t),\ \Theta_{22}(t)=\frac{1}{\langle{k}\rangle_{22}}
\sum\limits_{i=0}^{n_{21}}\sum\limits_{j=0}^{n_{22}}\sum\limits_{k=0}^{n_{23}}jP_B(i,j,k)\rho_{i,j,k}^B(t), $$

$$\Theta_{23}(t)=\frac{1}{\langle{k}\rangle_{23}}
\sum\limits_{i=0}^{n_{21}}\sum\limits_{j=0}^{n_{22}}\sum\limits_{k=0}^{n_{23}}kP_B(i,j,k)\rho_{i,j,k}^B(t),\ \Theta_{31}(t)=\frac{1}{\langle{k}\rangle_{31}}
\sum\limits_{i=0}^{n_{31}}\sum\limits_{j=0}^{n_{32}}\sum\limits_{k=0}^{n_{33}}iP_C(i,j,k)\rho_{i,j,k}^C(t), $$

$$\Theta_{33}(t)=\frac{1}{\langle{k}\rangle_{33}}
\sum\limits_{i=0}^{n_{31}}\sum\limits_{j=0}^{n_{32}}\sum\limits_{k=0}^{n_{33}}kP_C(i,j,k)\rho_{i,j,k}^C(t). $$

In the model, these special cases should be considered. (1) Isolated nodes as $(0,0,0)$-degree are not involved in disease spreading, and if the nodes are infected at initial state, they will eventually recover at rate $\mu_1$, $\mu_2$, and $\mu_3$; (2) If $n_{12}=n_{13}=n_{21}=n_{22}=n_{23}=n_{31}=n_{32}=n_{33}=0$, then the model has only one subnetwork, and the model becomes the classical SIS network model; (3) If $n_{12}=n_{21}=0$ (or $n_{23}=n_{32}=0$ or $n_{31}=n_{13}=0$), then a one-way circular-coupled network becomes a one-way string-coupled network; (4) If $n_{13}=n_{31}=n_{23}=n_{32}=n_{33}=0$ (or $n_{11}=n_{12}=n_{21}=n_{13}=n_{31}=0$ or $n_{12}=n_{21}=n_{22}=n_{23}=n_{32}=0$), then the model becomes one-way-coupled networks. \

An important epidemiological parameter is the basic reproduction number $R_0$, which is defined as the expected number of secondary infections produced by a single infective individual in a completely susceptible population. The basic reproduction number $R_0=\frac{\lambda}{\lambda_c}$ denotes the relationship between $R_0$ and the epidemic threshold $\lambda_c$ in physics.\

Hence, we will calculate the basic reproduction number $R_0$. Let $\rho_{0,0,0}^A=y_1, \rho_{0,0,1}^A=y_2, \cdots, \rho_{n_{11},n_{12},n_{13}}^A=y_{(n_{11}+1)(n_{12}+1)(n_{13}+1)}, \rho_{0,0,0}^B=y_{(n_{11}+1)(n_{12}+1)(n_{13}+1)+1}, \cdots, \rho_{n_{21},n_{22},n_{23}}^B=y_{(n_{11}+1)(n_{12}+1)(n_{13}+1)+(n_{21}+1)(n_{22}+1)(n_{23}+1)}, \\ \rho_{0,0,0}^C=y_{(n_{11}+1)(n_{12}+1)(n_{13}+1)+(n_{21}+1)(n_{22}+1)(n_{23}+1)+1}, \cdots, \rho_{n_{31},n_{32},n_{33}}^C=y_N,$ where
$N=[(n_{11}+1)(n_{12}+1)(n_{13}+1)+(n_{21}+1)(n_{22}+1)(n_{23}+1)+(n_{31}+1)(n_{32}+1)(n_{33}+1)]$. Then, $f=(f_1,f_2,\cdots,f_N)$ and $y=(y_1,y_2,\cdots,y_N)$ are both $N$-dimensional vectors, and $f$ denotes the right hand side of model $(\ref{e01})$. So, the model can be rewritten as the following:
\begin{equation}\label{e02}
\frac{dy(t)}{dt}=f(y(t)).
\end{equation}
It is easy to know the model has the disease-free equilibrium $E_0$ with $y_i=0$ for all $i=1,2,\cdots,N.$\

According to \cite{32}, the basic reproduction number can be computed by $R_0=\rho(FV^{-1})$, which represents the spectral radius of matrix $FV^{-1}$, $F$ is the rate of occurring new infections, and $V$ is the rate of transferring individuals outside of the original group. For the model, let $\mathscr{F}$ and $\mathscr{V}$ are

\begin{equation}\label{e03}     %¿ªÊ¼Êýѧ»·¾³
\mathscr{F}=\left( %×óÀ¨ºÅ
  \begin{array}{c}   %¸Ã¾ØÕóÒ»¹²3ÁУ¬Ã¿Ò»Áж¼¾ÓÖзÅÖÃ
    \lambda_{11}e(1-\rho_{e,f,g}^A(t))\Theta_{11}(t)+\lambda_{31}g(1-\rho_{e,f,g}^A(t))\Theta_{31}(t) \\  %µÚÒ»ÐÐÔªËØ
    \lambda_{12}l(1-\rho_{l,m,n}^B(t))\Theta_{12}(t)+\lambda_{22}m(1-\rho_{l,m,n}^B(t))\Theta_{22}(t)  \\  %µÚ¶þÐÐÔªËØ
    \lambda_{23}y(1-\rho_{x,y,z}^C(t))\Theta_{23}(t)+\lambda_{33}z(1-\rho_{x,y,z}^C(t))\Theta_{33}(t)   \\  %µÚÈýÐÐÔªËØ
  \end{array}
\right),  %ÓÒÀ¨ºÅ
\end{equation}

\begin{equation}\label{e04}     %¿ªÊ¼Êýѧ»·¾³
\mathscr{V}=\left( %×óÀ¨ºÅ
  \begin{array}{c}   %¸Ã¾ØÕóÒ»¹²3ÁУ¬Ã¿Ò»Áж¼¾ÓÖзÅÖÃ
    \mu_1\rho_{e,f,g}^A(t)  \\%µÚÒ»ÐÐÔªËØ
     \mu_2\rho_{l,m,n}^B(t) \\%µÚ¶þÐÐÔªËØ
      \mu_3\rho_{x,y,z}^C(t) \\%µÚÈýÐÐÔªËØ
  \end{array}
\right). %ÓÒÀ¨ºÅ
\end{equation}

By Ref.~\cite{32}, the matrix of new occurring infection rate $F=(\frac{\partial\mathscr{F}_i(E_0)}{\partial y_j})$, the matrix of individual transferring rate $V=(\frac{\partial\mathscr{V}_i(E_0)}{\partial y_j})$, where $i,j=0,1,\cdots,N$. Obviously, matrix $V$ is diagonal matrix, where $v_{ii}=\mu_1,\ i=1,\cdots,(n_{11}+1)(n_{12}+1)(n_{13}+1),\ v_{jj}=\mu_2,\ j=(n_{11}+1)(n_{12}+1)(n_{13}+1)+1,\cdots,(n_{11}+1)(n_{12}+1)(n_{13}+1)+(n_{21}+1)(n_{22}+1)(n_{23}+1),\ v_{kk}=\mu_3,\ k=(n_{11}+1)(n_{12}+1)(n_{13}+1)+(n_{21}+1)(n_{22}+1)(n_{23}+1)+1,\cdots,N$. Hence, we assume the joint degree distributions are independent, then
\begin{equation*}
     P_X(i,j,k)=P_X(i,\cdot,\cdot)P_X(\cdot,j,\cdot)P_X(\cdot,\cdot,k),\ X=A,B,C.
\end{equation*}

By means of similar transformations, the next-generation matrix $\Gamma=FV^{-1}$ can be rewritten as follows:
\begin{equation}\label{e05}     %¿ªÊ¼Êýѧ»·¾³
\Gamma=\left( %×óÀ¨ºÅ
  \begin{array}{cccccc}   %¸Ã¾ØÕóÒ»¹²6ÁУ¬Ã¿Ò»Áж¼¾ÓÖзÅÖÃ
    \frac{\lambda_{11}\langle{k^2}\rangle_{11}}{\mu_1\langle{k}\rangle_{11}} & \frac{\lambda_{11}}{\mu_1}\langle{k}\rangle_{13} & 0 & 0 & 0 & 0 \\% µÚÒ»ÐÐÔªËØ
     0 & 0 & 0 & 0 & \frac{\lambda_{31}}{\mu_3}\langle{k}\rangle_{32} & \frac{\lambda_{31}}{\mu_3}\langle{k}\rangle_{33} \\%µÚ¶þÐÐÔªËØ
      \frac{\lambda_{12}}{\mu_1}\langle{k}\rangle_{11} & \frac{\lambda_{12}}{\mu_1}\langle{k}\rangle_{13} & 0 & 0 & 0 & 0 \\% µÚÈýÐÐÔªËØ
      0 & 0 & \frac{\lambda_{22}}{\mu_2}\langle{k}\rangle_{21} & \frac{\lambda_{22}\langle{k^2}\rangle_{22}}{\mu_2\langle{k}\rangle_{22}} & 0 & 0 \\%µÚËÄÐÐÔªËØ
      0 & 0 & \frac{\lambda_{23}}{\mu_2}\langle{k}\rangle_{21} & \frac{\lambda_{23}}{\mu_2}\langle{k}\rangle_{22} & 0 & 0 \\%µÚÎåÐÐÔªËØ
      0 & 0 & 0 & 0 & \frac{\lambda_{33}}{\mu_3}\langle{k}\rangle_{32} & \frac{\lambda_{33}\langle{k^2}\rangle_{33}}{\mu_3\langle{k}\rangle_{33}} \\%µÚÁùÐÐÔªËØ
  \end{array}
\right). %ÓÒÀ¨ºÅ
\end{equation}

The basic reproduction number of model is $R_0=\rho(\Gamma)$, where $\rho(\Gamma)$ is the spectral radius of matrix $\Gamma$.

From the Theorem $2$ of Ref.~\cite{32}, we can obtain the following result.

\begin{theorem}\label{thm2.1}
For the disease transmission model $(\ref{e01})$, if $R_0<1$, then the disease-free equilibrium $E_0=(0,0,\cdots,0)$ is locally asymptotically stable; while if $R_0>1$, $E_0$ becomes unstable.
\end{theorem}

Now, we consider the global stability of the model $(\ref{e01})$.

\begin{theorem}\label{thm2.2}
 For model $(\ref{e01})$, $\Omega\triangleq\{y=\{y_1,y_2,\cdots,y_N\}, 0\leq y_i\leq 1, i=1,2,\cdots,N\}$ is a positive invariant set.
\end{theorem}
\noindent \textbf{Proof.} \ In order to prove the set $\Omega$ is positive invariant set. By using contradiction method, we prove that if initial value $y(0)\in\Omega$, then $y_i(t)\geq0$ for all $t>0, i=1,2,\cdots,N$. Otherwise, there are a $l_0\in\{1,2,\cdots,N\}$ and $t_0>0$, such that $y_{l_0}(t_0)=0$. Without loss of generality, we assume $y_{l_0}=\rho_{i_0,j_0,k_0}^A$ and $t^*=inf\{t>0,\rho_{i_0,j_0,k_0}^A(t)=0\}$, by the definition of $t^*$, we have $d\rho_{i_0,j_0,k_0}^A(t^*)/dt\leq0$.

However, by the model (\ref{e01}), it is easy to get $d\rho_{i_0,j_0,k_0}^A(t^*)/dt=\lambda_{11}i_0\Theta_{11}(t^*)+\lambda_{31}k_0\Theta_{21}(t^*)>0$, which is a contradiction. Because of $S_{i,j,k}^X(t)=1-\rho_{i,j,k}^X(t)$ for $X=A,B,C$, similarly, we can also verify that $S_{i,j,k}^X(t)>0$ for all $t>0$. The proof is finished. \hfill  $\Box$

Before discuss the global stability of the disease-free equilibrium and endemic equilibrium, we first introduce the Corollary $3.2$ in \cite{33} as the Lemma $1$ as follow:

\begin{lemma}[\cite{33}]\label{lem2.1}
 For the model $(\ref{e02})$, let $f: R_+^n\rightarrow R^n$ be a continuously differentiable map. Assume that

 (1) $f$ is cooperative on $R_+^n$ and $Df(y)=(\partial f_i/\partial y_j)_{1\leq i,j\leq N}$ is irreducible for every $y\in R_+^n$;\

(2) $f(0)=0$ and $f_i(y)\geq0$ for all $y\in R_+^n$ with $y_i=0,\
i=1,2,\cdots,N$;\

(3) $f$ is strictly sublinear on $y\in R_+^n$, i.e., for any $\alpha\in(0,1)$ and any $y\gg0,\ f(\alpha y)>\alpha f(y)$.\

(a) If $s(Df(0))\leq0$, then $y=0$ is globally asymptotically stable with respect to $y\in R_+^n$;\

(b) If $s(Df(0))>0$, then either\

\ (i) for any $y\in R_+^n\setminus\{0\},\ \lim_{t\rightarrow\infty}|\varphi(t,y)|=+\infty$, or alternatively,\

\ (ii) the model admits a unique positive steady state $y^*\gg0$ and $y=y^*$ is globally asymptotically stable with respect to $R_+^n\setminus\{0\}$.
\end{lemma}

Next, we give the following Theorem to show the stability analysis.\

\begin{theorem}\label{thm2.3 }
If $R_0<1$, then the disease-free equilibrium $E_0$ is globally asymptotically stable in $\Omega$; but if $R_0>1$, then model $(\ref{e01})$ admits a unique endemic equilibrium $E^*$, which is globally asymptotically stable in $\Omega-\{\mathbf{0}\}$.
\end{theorem}
\noindent \textbf{Proof.} \ It is easy to show the model $(\ref{e01})$ satisfies to the condition $(2)$ of the above Lemma, the function $f: \Omega\rightarrow R^n$ is continuously differentiable, and $f$ is cooperative. Further, we know that $Df(y)=(\partial f_i/\partial y_i)_{1\leq i,j\leq n}$ is irreducible for $y\in\Omega$. Moreover, for any $\varepsilon\in(0,1)$ and $y\in\Omega$, $f_i(\varepsilon y)\geq\varepsilon f_i(y), i=1,2,\cdots,N$, which implies that $f$ is strictly sublinear in $\Omega$.

By Lemma $1$, if $R_0\leq1 (s(Df(0))\leq0)$, then the disease-free equilibrium $E_0$ is globally asymptotically stable in $\Omega$, if $R_0\geq1 (s(Df(0))>0)$, by Theorem $2$, the case (i) is impossible, so the model $(\ref{e01})$ admits a unique endemic equilibrium $E^*$, which is globally asymptotically stable in $\Omega-\{\mathbf{0}\}$. \hfill  $\Box$

In the model, the basic reproduction $R_0$ cannot be explicitly expressed, but by using the Perron-Frobenius theorem, we know $R_0$ increase with the increase of infection rates $\lambda$, infection periods $1/\mu$, average degrees $\langle{k}\rangle$ and degree ratio $\langle{k^2}\rangle/\langle{k}\rangle$. So, we can estimate the range of $R_0$ as follows
$$\min\{r_1,r_2,\cdots,r_6\}\leq R_0\leq\max\{r_1,r_2,\cdots,r_6\},$$
where $r_i$ is the sum of the elements in the $i$-th row (or the $i$-th column) of matrix (\ref{e05}).

\section{One-way circular-coupled network with infective media}

In this section, we consider a one-way circular-coupled network with infective media. The interaction model contains three populations network $A$, $B$ and $C$, and three vectors $a$, $b$ and $c$. Assume that there is no direct contact among in network $A$, $B$ and $C$, the infectious disease transmission through the vectors. The direction of propagation is one-way and becomes a cycle network. A schematic diagram is presented in Fig.~\ref{e002}.
\begin{figure}[]
\begin{center}
\includegraphics[width=4cm,height=4cm]{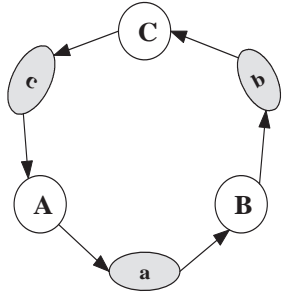}
\caption{A one-way circular-coupled network with three subnetworks and three vectors.}\label{e002}
\end{center}
\end{figure}

The disease spreading process is described as follows. At each time step, a susceptible node in subnetwork $A$ (or $B$ or $C$) is infected with probability $\lambda_1 $(or $\lambda_2$ or $\lambda_3$) if it is connected to an infected in the same subnetwork. A susceptible node in subnetwork $A$ also infected with probability $r_{32}$ if it is contacts with infected vector $c$, and the vectors $b$ and $c$ have no effect on the node in subnetwork $A$. Similarly, a susceptible node in subnetwork $B$ (or $C$) is infected with probability $r_{12}$ (or $r_{22}$) if it is contacts with infected vector $a$ (or $b$). On the other hand, each susceptible vector $a$ (or $b$ or $c$) is infected with probability $r_{11}$ (or $r_{21}$  or $r_{31}$) through contacts with infected individuals in subnetwork $A$ (or $B$ or $C$). All infected individuals and infected vectors may be recovered and become susceptible again. Without loss of generality, we set the recovery rates of all subnetworks and vectors be one. So, the three networks and the three vectors are SIS epidemic model.

Hence, let $\rho_k(t)$, $\eta_l(t)$ and $\xi_m(t)$ be the densities of infected nodes with degree $k$, $l$ and $m$ at time $t$ in subnetworks $A$, $B$ and $C$, respectively. Let $\vartheta_1$, $\vartheta_2$ and $\vartheta_3$ be the densities of an infective vector a, b and c at time $t$. Here, $k=1,2,\cdots,n$, $l=1,2,\cdots,p$ and $m=1,2,\cdots,q$, where $n$, $p$ and $q$ is the maximum connectivity number in subnetworks A, B and C, respectively.

Then, the dynamical mean-field reaction rate equations of the whole model can be described as:
\begin{equation}\label{e06}
\left\{\begin{array}{ll}
\dot{\rho}_{k}(t)=-\rho_k(t)+(1-\rho_k(t))(\lambda_1k\Theta_1(t)+r_{32}\vartheta_3(t)),  \\
\dot{\vartheta}_{1}(t)=-\vartheta_1(t)+r_{11}\rho(t)(1-\vartheta_1(t)), \\
\dot{\eta}_{l}(t)=-\eta_l(t)+(1-\eta_l(t))(\lambda_2l\Theta_2(t)+r_{12}\vartheta_1(t)), \\
\dot{\vartheta}_{2}(t)=-\vartheta_2(t)+r_{21}\eta(t)(1-\vartheta_2(t)), \\
\dot{\xi}_{m}(t)=-\xi_m(t)+(1-\xi_m(t))(\lambda_3m\Theta_3(t)+r_{22}\vartheta_2(t)), \\
\dot{\vartheta}_{3}(t)=-\vartheta_3(t)+r_{31}\xi(t)(1-\vartheta_3(t)),
\end{array}\right.
\end{equation}
where $\rho(t)=\sum\limits_{k=1}^{n}P_1(k)\rho_k(t)$, $\eta(t)=\sum\limits_{l=1}^{p}P_2(l)\eta_l(t)$ and $\xi(t)=\sum\limits_{m=1}^{q}P_3(m)\xi_m(t)$ denote the global infection density in subnetworks $A$, $B$ and $C$, in which $P_1(k)$, $P_2(l)$ and $P_3(m)$ are the degree distributions of three subnetworks $A$, $B$ and $C$, respectively. Since the degrees of the network are uncorrelated, $\Theta_1(t)$ ($\Theta_2(t)$ or $\Theta_3(t)$) represents the probability of a randomly chosen link pointing to an infected node in subnetwork $A$ ($B$ or $C$), they can be written as
$$\Theta_1(t)=\frac{1}{\langle{k}\rangle}\sum\limits_{k=1}^{n}kP_1(k)\rho_k(t),\ \Theta_2(t)=\frac{1}{\langle{l}\rangle}\sum\limits_{l=1}^{p}lP_2(l)\eta_l(t),\ \Theta_3(t)=\frac{1}{\langle{m}\rangle}\sum\limits_{m=1}^{q}mP_3(m)\xi_m(t),$$
in which $\langle{k}\rangle=\sum\limits_{k=1}^{n}kP_1(k)$, $\langle{l}\rangle=\sum\limits_{l=1}^{p}lP_2(l)$ and $\langle{m}\rangle=\sum\limits_{m=1}^{q}mP_3(m)$ are the average degrees of subnetworks $A$, $B$ and $C$, respectively.

Let $\rho_k(t)=y_k(t)$ for $k=1,2,\cdots,n$, $\eta_l(t)=y_{n+l}(t)$ for $l=1,2,\cdots,p$, $\xi_m(t)=y_{n+p+m}(t)$ for $m=1,2,\cdots,q$, $\vartheta_1(t)=y_{n+p+q+1}(t)$, $\vartheta_2(t)=y_{n+p+q+2}(t)$, $\vartheta_3(t)=y_{n+p+q+3}(t)$, and
$$\tilde{\Omega}=\{(y_1,y_2,\cdots,y_{n+p+q+3}),\ 0\leq y_i\leq1,\ i=1,2,\cdots,n+p+q+3\}.$$

According to Ref.~\cite{32}, the basic reproduction number can be computed by $R_0=\rho(FV^{-1})$. For the model (\ref{e06}), let $\mathscr{F}$ and $\mathscr{V}$ are

\begin{equation}\label{e07}     %¿ªÊ¼Êýѧ»·¾³
\mathscr{F}=\left( %×óÀ¨ºÅ
  \begin{array}{c}   %¸Ã¾ØÕóÒ»¹²3ÁУ¬Ã¿Ò»Áж¼¾ÓÖзÅÖÃ
              (1-\rho_k(t))(\lambda_1k\Theta_1(t)+r_{32}\vartheta_3(t))  \\%µÚÒ»ÐÐÔªËØ
              (1-\eta_l(t))(\lambda_2l\Theta_2(t)+r_{12}\vartheta_1(t))  \\%µÚ¶þÐÐÔªËØ
              (1-\xi_m(t))(\lambda_3m\Theta_3(t)+r_{22}\vartheta_2(t))  \\% µÚÈýÐÐÔªËØ
              r_{11}\rho(t)(1-\vartheta_1(t))   \\
              r_{21}\eta(t)(1-\vartheta_2(t))  \\
              r_{31}\xi(t)(1-\vartheta_3(t))  \\
  \end{array}
\right),  %ÓÒÀ¨ºÅ
\end{equation}

\begin{equation}\label{e08}     %¿ªÊ¼Êýѧ»·¾³
\mathscr{V}=\left( %×óÀ¨ºÅ
  \begin{array}{c}   %¸Ã¾ØÕóÒ»¹²3ÁУ¬Ã¿Ò»Áж¼¾ÓÖзÅÖÃ
             \rho_k(t)  \\
             \eta_l(t) \\
             \xi_m(t)  \\
             \vartheta_1(t)  \\
             \vartheta_2(t)  \\
             \vartheta_3(t) \\
  \end{array}
\right). %ÓÒÀ¨ºÅ
\end{equation}

Then, $V$ is an $(n+p+q+3)\times(n+p+q+3)$ identity matrix, and
\begin{equation}\label{e09}     %¿ªÊ¼Êýѧ»·¾³
F=\left( %×óÀ¨ºÅ
  \begin{array}{ccccccccccccccc}   %¸Ã¾ØÕóÒ»¹²6ÁУ¬Ã¿Ò»Áж¼¾ÓÖзÅÖÃ
     P_1 & P_2 & \cdots & P_n & 0 & 0 & \cdots & 0 & 0 & 0 & \cdots & 0 & 0 & 0 & r_{32} \\% µÚÒ»ÐÐÔªËØ
     2P_1 & 2P_2 & \cdots & 2P_n & 0 & 0 & \cdots & 0 & 0 & 0 & \cdots & 0 & 0 & 0 & r_{32} \\%µÚ¶þÐÐÔªËØ
     \cdots & \cdots & \cdots & \cdots & \cdots & \cdots & \cdots & \cdots & \cdots & \cdots & \cdots & \cdots & \cdots & \cdots & \cdots\\% µÚÈýÐÐÔªËØ
     nP_1 & nP_2 & \cdots & nP_n & 0 & 0 & \cdots & 0 & 0 & 0 & \cdots & 0 & 0 & 0 & r_{32}  \\%µÚËÄÐÐÔªËØ
      0 & 0 & \cdots & 0 & Q_1 & Q_2 & \cdots & Q_p & 0 & 0 & \cdots & 0 & r_{12} & 0 & 0 \\%µÚÎåÐÐÔªËØ
      0 & 0 & \cdots & 0 & 2Q_1 & 2Q_2 & \cdots & 2Q_p & 0 & 0 & \cdots & 0 & r_{12} & 0 & 0 \\%µÚÁùÐÐÔªËØ
      \cdots & \cdots & \cdots & \cdots & \cdots & \cdots & \cdots & \cdots & \cdots & \cdots & \cdots & \cdots & \cdots & \cdots & \cdots\\% µÚÆßÐÐÔªËØ
      0 & 0 & \cdots & 0 & pQ_1 & pQ_2 & \cdots & pQ_p & 0 & 0 & \cdots & 0 & r_{12} & 0 & 0 \\%µÚ°ËÐÐÔªËØ
      0 & 0 & \cdots & 0 & 0 & 0 & \cdots & 0 & R_1 & R_2 & \cdots & R_q & 0 & r_{22} & 0 \\%µÚ¾ÅÐÐÔªËØ
      0 & 0 & \cdots & 0 & 0 & 0 & \cdots & 0 & 2R_1 & 2R_2 & \cdots & 2R_q & 0 & r_{22} & 0 \\%µÚÊ®ÐÐÔªËØ
      \cdots & \cdots & \cdots & \cdots & \cdots & \cdots & \cdots & \cdots & \cdots & \cdots & \cdots & \cdots & \cdots & \cdots & \cdots\\% µÚʮһÐÐÔªËØ
      0 & 0 & \cdots & 0 & 0 & 0 & \cdots & 0 & qR_1 & qR_2 & \cdots & qR_q & 0 & r_{22} & 0 \\%µÚÊ®¶þÐÐÔªËØ
      A_1 & A_2 & \cdots & A_n & 0 & 0 & \cdots & 0 & 0 & 0 & \cdots & 0 & 0 & 0 & 0 \\% µÚÊ®ÈýÐÐÔªËØ
      0 & 0 & \cdots & 0 & B_1 & B_2 & \cdots & B_p & 0 & 0 & \cdots & 0 & 0 & 0 & 0 \\% µÚÊ®ËÄÐÐÔªËØ
      0 & 0 & \cdots & 0 & 0 & 0 & \cdots & 0 & C_1 & C_2 & \cdots & C_q & 0 & 0 & 0 \\% µÚÊ®ÎåÐÐÔªËØ
  \end{array}
\right). %ÓÒÀ¨ºÅ
\end{equation}
where $P_k=\frac{1}{\langle{k}\rangle}\lambda_1kP_1(k)$, $Q_l=\frac{1}{\langle{l}\rangle}\lambda_2lP_2(l)$, $R_m=\frac{1}{\langle{m}\rangle}\lambda_3mP_3(m)$, $A_k=r_{11}P_1(k)$, $B_l=r_{21}P_2(l)$, $C_m=r_{31}P_3(m)$.

Hence, the next-generation matrix $\Gamma=FV^{-1}=F$, by the similarity transformation and ignoring some zero blocks, the matrix $\Gamma$ is transformed into
\begin{equation}\label{e10}     %¿ªÊ¼Êýѧ»·¾³
\tilde{\Gamma}=\left( %×óÀ¨ºÅ
  \begin{array}{ccccccccc}   %¸Ã¾ØÕóÒ»¹²6ÁУ¬Ã¿Ò»Áж¼¾ÓÖзÅÖÃ
      \lambda_1\frac{\langle{k^2}\rangle}{\langle{k}\rangle} & \lambda_1 r_{32} & 0 & 0 & 0 & 0 & 0 & 0 & 0  \\% µÚÒ»ÐÐÔªËØ
      0 & 0 & 0 & 0 & 0 & 0 & 0 & 0 & 1 \\%µÚ¶þÐÐÔªËØ
      0 & 0 & \lambda_2\frac{\langle{l^2}\rangle}{\langle{l}\rangle} & \lambda_2 r_{12} & 0 & 0 & 0 & 0 & 0 \\% µÚÈýÐÐÔªËØ
      0 & 0 & 0 & 0 & 0 & 0 & 1 & 0 & 0 \\%µÚËÄÐÐÔªËØ
      0 & 0 & 0 & 0 & \lambda_3\frac{\langle{m^2}\rangle}{\langle{m}\rangle} & \lambda_3 r_{22} & 0 & 0 & 0 \\%µÚÎåÐÐÔªËØ
      0 & 0 & 0 & 0 & 0 & 0 & 0 & 1 & 0 \\%µÚÁùÐÐÔªËØ
      r_{11}\langle{k}\rangle & r_{11}r_{32} & 0 & 0 & 0 & 0 & 0 & 0 & 0\\% µÚÆßÐÐÔªËØ
      0 & 0 & r_{21}\langle{l}\rangle &r_{21}r_{12} & 0 & 0 & 0 & 0 & 0 \\% µÚ°ËÐÐÔªËØ
      0 & 0 & 0 & 0 & r_{31}\langle{m}\rangle &r_{31}r_{22} & 0 & 0 & 0 \\% µÚ¾ÅÐÐÔªËØ
  \end{array}
\right). %ÓÒÀ¨ºÅ
\end{equation}

So, the basic reproduction number of model (\ref{e06}) is $R_0=\rho(\tilde{\Gamma})$. Since $\tilde{\Gamma}$ is a nonnegative definite matrix, it follows that $\tilde{\Gamma}$ has a positive eigenvalue equaling to $R_0$.

According to Theorem 2 in Ref.~\cite{32}, we can obtain the following result.

\begin{theorem}\label{thm3.1 }
 If $R_0<1$, then the disease-free equilibrium $E_0=(0,0,\cdots,0)$ is locally asymptotically stable, while if $R_0>1$, then $E_0$ becomes unstable.
\end{theorem}

Next, we analysis the global behavior of system (\ref{e06}). To guarantee the positivity and boundedness of the system, the following Lemma is needed.

\begin{lemma}\label{lem3.1}
The set $\tilde{\Omega}$ is positive invariant for the model (\ref{e06}).
\end{lemma}
\noindent \textbf{Proof.} \ We will prove that if $y(0)\in\tilde{\Omega}$, then $y(t)\in\tilde{\Omega}$ for all $t>0$. Let
$$\partial\tilde{\Omega}_1=\{y\in\tilde{\Omega}|y_i=0\ for\ some\ i\},\ \partial\tilde{\Omega}_2=\{y\in\tilde{\Omega}|y_i=1\ for\ some\ i\},$$
where $i=1,2,\cdots,n+p+q+3$, and the 'outer normals' be denotes as $\delta_i^1=(\overbrace{0,\cdots,0,-1}^i,0,\cdots,0)$ and $\delta_i^2=(\overbrace{0,\cdots,0,1}^i,0,\cdots,0)$.

For an arbitrary set $\Delta$, Ref.~\cite{34} had proved that $\Delta$ is invariant for $dx/dt=f(x)$, if at each point $y$ on the boundary of $\Delta$, the vector $f(y)$ is tangent or pointing into the set. It is easy to apply this results here, we can get
\begin{eqnarray*}
  \frac{dy}{dt}\mid_{y_i=0}\cdot \delta_i^1 &=& -(\frac{\lambda_1k}{\langle{k}\rangle}\sum\limits_{k\neq i}kP_1(k)\rho_k+r_{32}\vartheta_3)\leq0,\ i=1,\cdots,n, \\
  \frac{dy}{dt}\mid_{y_j=0}\cdot \delta_j^1 &=& -(\frac{\lambda_2l}{\langle{l}\rangle}\sum\limits_{l\neq j}lP_2(l)\eta_l+r_{12}\vartheta_1)\leq0,\ j=n+1,\cdots,n+p, \\
  \frac{dy}{dt}\mid_{y_s=0}\cdot \delta_s^1 &=& -(\frac{\lambda_3m}{\langle{m}\rangle}\sum\limits_{m\neq s}mP_3(m)\xi_m+r_{22}\vartheta_2)\leq0,\ s=n+p+1,\cdots,n+p+q, \\
  \frac{dy}{dt}\mid_{y_g=0}\cdot \delta_g^1 &=& -r_{11}\sum\limits_{k=1}^nP_1(k)\rho_k\leq0,\ g=n+p+q+1, \\
  \frac{dy}{dt}\mid_{y_h=0}\cdot \delta_h^1 &=& -r_{21}\sum\limits_{l=1}^pP_2(l)\eta_l\leq0,\ h=n+p+q+2, \\
  \frac{dy}{dt}\mid_{y_r=0}\cdot \delta_r^1 &=& -r_{31}\sum\limits_{m=1}^qP_3(m)\xi_m\leq0,\ r=n+p+q+3,  \\
  \frac{dy}{dt}\mid_{y_u=1}\cdot \delta_u^2 &=& -y_u\leq0,\ u=1,2,\cdots,n+p+q+3.
\end{eqnarray*}

Therefore, any solution that starts in $y\in\partial\tilde{\Omega}_1\bigcup\partial\tilde{\Omega}_2$ will stays inside $\tilde{\Omega}$.  \hfill  $\Box$

Let $y=(y_1,y_2,\cdots,y_{n+p+q+3})\in\tilde{\Omega}$ and $L(y)$ be a column vector, then, Eq.(\ref{e06}) can be rewritten as a compact vector form:
\begin{equation}\label{e11}
 \frac{dy}{dt}=Ay+L(y),
\end{equation}
where $Ay$ is the linear part of $y$, and $L(y)$ is the nonlinear part of $y$. For $L(y)$, the $k$-th component is $-y_k[k\lambda_1\Theta_1(y)+r_{32}y_{n+p+q+3}]$, $k=1,2,\cdots,n$, the $l$-th component is $-y_l[l\lambda_2\Theta_2(y)+r_{12}y_{n+p+q+1}]$, $l=n+1,n+2,\cdots,n+p$, the $m$-th component is $-y_m[m\lambda_3\Theta_3(y)+r_{22}y_{n+p+q+2}]$, $m=n+p+1,n+p+2,\cdots,n+p+q$, the $(n+p+q+1)$-th component is $-y_{n+p+q+1}r_{11}X$, the $(n+p+q+2)$-th component is $-y_{n+p+q+2}r_{21}Y$, and the $(n+p+q+3)$-th component is $-y_{n+p+q+3}r_{31}Z$, in which $\Theta_1(y)=\langle{k}\rangle^{-1}\sum_{k=1}^{n}kP_1(k)y_k(t)$, $\Theta_2(y)=\langle{l}\rangle^{-1}\sum_{l=1}^{p}lP_2(l)y_{n+l}(t)$, $\Theta_3(y)=\langle{m}\rangle^{-1}\sum_{m=1}^{q}mP_3(m)y_{n+p+m}(t)$, $X(t)=\sum_{k=1}^{n}P_1(k)y_k(t)$,
$Y(t)=\sum_{l=1}^{p}P_2(l)y_{n+l}(t)$, and $Z(t)=\sum_{m=1}^{q}P_3(m)y_{n+p+m}(t)$.

Denote $s(A)=max_{1\leq i\leq n+p+q+3}Re\ \lambda_i$, where $\lambda_i$ for $i=1,2,\cdots,n+p+q+3$ is the eigenvalue of $A$ and $Re\ \lambda_i$ represents the real part of $\lambda_i$.

\begin{remark}\label{rem1}
$s(A)<0\Longleftrightarrow R_0<1$, $s(A)>0\Longleftrightarrow R_0>1$.
\end{remark}

To study the global stability of the model, the following Lemma is also needed.

\begin{lemma}[\cite{35}]\label{lem3.2}
Consider the system
\begin{equation}\label{e12}
 \frac{dy}{dt}=Ay+L(y),
\end{equation}
where $A$ is an $n\times n$ matrix and $L(y)$ is continuously differentiable in a region $D\in R^n$. Assume that

(1) the compact convex set $C\subset D$ is positively invariant with respect to (\ref{e12}), and $0\in C$,

(2) $\lim_{y\rightarrow0}\|L(y)\|/\|y\|=0$,

(3) there exist $r > 0$ and a (real) eigenvector $\omega$ of $A^T$ such that $(\omega\cdot y)\geq r\|y\|$ for all $y\in C$,

(4) $(\omega\cdot L(y))\leq0$ for all $y\in C$,

(5) $y=0$ is the largest positively invariant set contained in $H=\{y\in C|(\omega\cdot L(y))=0\}$. Then, either $y=0$ is globally asymptotically stable in $C$, or for any $y_0\in C-\{\mathbf{0}\}$ the solution $\varphi(t,y_0)$ of (\ref{e12}) satisfies $\liminf_{t\rightarrow\infty}\|\varphi(t,y_0)\|\geq m$, where $m>0$ is independent of the initial value $y_0$. Moreover, there exists a constant solution of (\ref{e12}), $y=y^*,\ y^*\in C-\{\mathbf{0}\}$.
\end{lemma}
Then, we have the following result:
\begin{theorem}\label{thm3.2}
 For the disease transmission model ($\ref{e06})$, if $R_0<1$, then the disease-free equilibrium $E_0$ is globally asymptotically stable in $\tilde{\Omega}$, if $R_0>1$, there exists a unique positive constant solution $y^*\in \tilde{\Omega}-\{\mathbf{0}\}$.
\end{theorem}
\noindent \textbf{Proof.} \ By using Lemma $3$, we will confirm that for system $(\ref{e11})$ meet all hypotheses of Lemma $3$. Condition (1) is satisfied by letting $C=\tilde{\Omega}$. Conditions (2) and (4) are clearly satisfied. For condition (3), notice that $A^T$ is irreducible and $a_{ij}\geq0$ whenever $i\neq j$, then there exists an eigenvector $x=(x_1,x_2,\cdots,x_g)$ of $A^T$ and the corresponding eigenvalue is $s(A^T),\ g=n+p+q+3$. Let $x_0=\min_ix_i$, for $y\in\tilde{\Omega}$, we obtain $(x\cdot y)\geq x_0\sum_{i=1}^g y_i\geq x_0(\sum_{i=1}^gy_i^2)^{\frac{1}{2}}$. Hence, condition (3) is reached by letting $r=x_0$. To verify condition (5), we set $G=\{y\in\tilde{\Omega}|(\omega\cdot L(y))=0\}$. If $y\in G$, then
\begin{eqnarray*}
  && \sum\limits_{i=1}^{n}x_iy_i[\lambda_1i\Theta_1(y)+r_{32}y_{n+p+q+3}]+\sum\limits_{i=1}^{p}x_{n+i}y_{n+i}[\lambda_2i\Theta_2(y)+r_{12}y_{n+p+q+1}] \\
  && +\sum\limits_{i=1}^{q}x_{n+p+i}y_{n+p+i}[\lambda_3i\Theta_3(y)+r_{22}y_{n+p+q+2}]+x_{n+p+q+1}y_{n+p+q+1}\cdot r_{11}X \\
  && +x_{n+p+q+2}y_{n+p+q+2}\cdot r_{21}Y+x_{n+p+q+3}y_{n+p+q+3}\cdot r_{31}Z=0.
\end{eqnarray*}

Because of each term of the above sum is nonnegative, here we have $y=0$ for all $i=1,2,\cdots,g$. Therefore, the only invariant set with
respect to (\ref{e11}) contained in $G$ is $y=0$, so condition (5) is satisfied. Hereby, all hypotheses of Lemma $3$ are satisfied.

Next, we will prove that if $R_0>1$, there is only one constant solution $y=y^*$ in $\tilde{\Omega}-\{\mathbf{0}\}$. Assume that $y=y^*$ and $y=z^*>0$ are two constant solutions of (\ref{e11}) (also (\ref{e06})) in $\tilde{\Omega}-\{\mathbf{0}\}$. If $y^*\neq z^*$, then there exists at least one $i_0,\ i_0=1,2,\cdots,n+p+q+3$, such that $y_{i_0}^*\neq z_{i_0}^*$, in which $y_{i_0}$ is the $i_0$-th component of the vector $y^*$. Without loss of generality, we assume $y_{i_0}^*>z_{i_0}^*$ and moreover $y_{i_0}^*/z_{i_0}^*\geq y_{i}^*/z_{i}^*$ for all $i=1,2,\cdots,n+p+q+3$. Then, we substitute them into $(\ref{e11})$ satisfies
\begin{equation*}
-y_{i_0}^*+(1-y_{i_0}^*)[\lambda_1\Theta_1(y^*)+r_{32}y_{n+p+q+3}^*]=-z_{i_0}^*+(1-z_{i_0}^*)[\lambda_1\Theta_1(z^*)+r_{32}z_{n+p+q+3}^*]=0.
\end{equation*}

After the equivalent reformulation, it follows that
\begin{equation*}
-z_{i_0}^*+(1-y_{i_0}^*)\frac{z_{i_0}^*}{y_{i_0}^*}[\lambda_1\Theta_1(y^*)+r_{32}y_{n+p+q+3}^*]=-z_{i_0}^*+(1-z_{i_0}^*)[\lambda_1\Theta_1(z^*)+r_{32}z_{n+p+q+3}^*]=0.
\end{equation*}

But $(z_{i_0}^*/y_{i_0}^*)y_i^*\leq z_i^*$ for all $i$ and $1-y_{i_0}^*<1-z_{i_0}^*$, thus from the above equality we obtain
\begin{equation*}
(1-y_{i_0}^*)\frac{z_{i_0}^*}{y_{i_0}^*}[\lambda_1\Theta_1(y^*)+r_{32}y_{n+p+q+3}^*]<(1-z_{i_0}^*)[\lambda_1\Theta_1(z^*)+r_{32}z_{n+p+q+3}^*].
\end{equation*}

This is a contradiction. Therefore, there is only one constant solution $y^*=(y_1^*,\cdots,y_{n+p+q+3}^*)$ in $\tilde{\Omega}-\{\mathbf{0}\}$. The proof is completed. \hfill  $\Box$

\begin{theorem}\label{thm3.3}
If $R_0>1$, then the unique endemic equilibrium $y^*=(y_1^*,y_2^*,\cdots,y_{n+p+q+3}^*)$ of model (\ref{e06}) is globally attractive in $\tilde{\Omega}-\{\mathbf{0}\}$.
\end{theorem}
\noindent \textbf{Proof.} \ Now, we define the following functions, $f:\ \tilde{\Omega}\rightarrow R$ and $F:\ \tilde{\Omega}\rightarrow R$, where $f(y)=\max_i(y_i/y_i^*)$ and $F(y)=\min_i(y_i/y_i^*)$ for $y\in\tilde{\Omega}$. $f(y)$ and $F(y)$ are continuous and their derivatives exists along solutions of (\ref{e11}). Let $y=y(t)$ be a solution of system (\ref{e11}). For a given $t_0$ and a sufficiently small $\epsilon>0$, we assume that $f(y(t))=y_{i_0}(t)/y_{i_0}^*$ for $t\in[t_0,t_0+\epsilon]$, $1\leq i_0\leq n+p+q+3$. Then,
$$f'|_{(11)}(y(t_0))=\frac{y'_{i_0}(t_0)}{y_{i_0}^*},\ t\in[t_0,t_0+\epsilon],$$
where $f'|_{(11)}=\limsup_{h\rightarrow0^+}\frac{f(y(t+h))-f(y(t))}{h}$.

From (\ref{e11}) we obtain
\begin{equation*}
 y_{i_0}^*\frac{y'_{i_0}(t_0)}{y_{i_0}(t_0)}=-y_{i_0}^*+(1-y_{i_0}(t_0))\frac{y_{i_0}^*}{y_{i_0}(t_0)}[\lambda_1\Theta_1(y^*)+r_{32}y_{n+p+q+3}^*].
\end{equation*}

According to the definition of $f(y(t))$, it follows that
$$\frac{y_{i_0}(t_0)}{y_{i_0}^*}\geq\frac{y_i(t_0)}{y_i^*},\ i=1,2,\cdots,n+p+q+3.$$

Then if $f(y(t_0))>1$, we can get
\begin{equation*}
 y_{i_0}^*\frac{y'_{i_0}(t_0)}{y_{i_0}(t_0)}<-y_{i_0}^*+(1-y_{i_0}(t_0))[\lambda_1\Theta_1(y^*)+r_{32}y_{n+p+q+3}^*]=0.
\end{equation*}
Since $y_{i_0}^*>0$ and $y_{i_0}(t_0)>0$, from the above inequality we obtain that $y'_{i_0}(t_0)<0$. So, $f(y(t_0))>1$ deduces $f'|_{(11)}(y(t_0))<0$.

Similarly, we can testify if $f(y(t_0))=1$, then $f'|_{(11)}(y(t_0))\leq0$. And if $f(y(t_0))<1$, we obtain $f'|_{(11)}(y(t_0))>0$. By the same method, it can be easily verified that if $F(y(t_0))>1$, then $F'|_{(11)}(y(t_0))<0$; if $F(y(t_0))=1$, then $F'|_{(11)}(y(t_0))\geq0$; if $F(y(t_0))<1$, then $F'|_{(11)}(y(t_0))>0$. Denote
$$U(y)=\max\{f(y)-1,0\},\ V(y)=\max\{1-F(y),0\}.$$

We know $U(y)$ and $V(y)$ are continuous and nonnegative for $y\in\tilde{\Omega}$. Note that, for any $t>0$, there exist
$$U'|_{(11)}(y(t))\leq0,\ V'|_{(11)}(y(t))\leq0.$$

Let
$$H_U=\{y\in\tilde{\Omega}|U'|_{(11)}(y(t))=0\},\ H_V=\{y\in\tilde{\Omega}|V'|_{(11)}(y(t))=0\}.$$
then $H_U=\{y|0\leq y_i\leq y_i^*\}$ and $H_V=\{y|y_i^*\leq y_i\leq1\}\cup\{\mathbf{0}\}$.

According to the LaSalle invariance principle, any solution of system (\ref{e11}) beginning in $\tilde{\Omega}$ will reach $H_U\cap H_V$, and $H_U\cap H_V=\{y_i^*\}\cup\{\mathbf{0}\}$. However, by Lemma $3$, if $y(t)\neq0$ and $R_0>1$, we have that $\liminf_{t\rightarrow\infty}\|y(t)\|\geq m>0$. Then, we get any solution of (\ref{e11}), and if $y(0)\in\tilde{\Omega}-\{\mathbf{0}\}$ satisfies $\lim_{t\rightarrow\infty}y(t)=y^*$, so $y=y^*$ is globally attractive in $\tilde{\Omega}-\{\mathbf{0}\}$. \hfill  $\Box$

\section{Numerical analysis}
In this section, we present some numerical examples on the SIS epidemic model to complement our theoretical results and further explore the transmission dynamics.
\subsection{One-way circular-coupled network}
In order to study how the interactions between subnetworks influence on the epidemic spread of the network, we present abundant numerical simulations on system $(\ref{e01})$. We set the joint degree distributions are independent, so we consider two kinds of networks, scale-free network and random network \cite{36}. The scale-free network is a heterogeneous network, in which the degree is $P(k)=k^{-\gamma}/\sum_kk^{-\gamma}$, the maximum degree $n\approx k_0\sqrt[\gamma-1]{N}$, where $N$ is the number of total nodes, $k_0$ is the minimum degree of the network. The random network is a homogeneous network, whose degree distribution is Poisson $P(k)=\lambda^ke^{-\lambda}/k!$, where constant $\lambda$ denotes the average degree. In the simulations, A, B and C denotes the contact pattern in subnetwork $A$, $B$ and $C$, respectively. And AB represent the contact pattern in subnetwork $A$ connecting $B$, similar to the meaning of BC and CA.  Let $$\rho^A(t)=\sum\limits_{i=0}^{n_{11}}\sum\limits_{j=0}^{n_{12}}\sum\limits_{k=0}^{n_{13}}P_A(i,j,k)\rho_{i,j,k}^A(t),\ \rho^B(t)=\sum\limits_{i=0}^{n_{21}}\sum\limits_{j=0}^{n_{22}}\sum\limits_{k=0}^{n_{23}}P_B(i,j,k)\rho_{i,j,k}^B(t),\ \rho^C(t)=\sum\limits_{i=0}^{n_{31}}\sum\limits_{j=0}^{n_{32}}\sum\limits_{k=0}^{n_{33}}P_C(i,j,k)\rho_{i,j,k}^C(t)$$ as the corresponding total infected densities.

\begin{figure}[]
\begin{center}
\includegraphics[width=10cm,height=5cm]{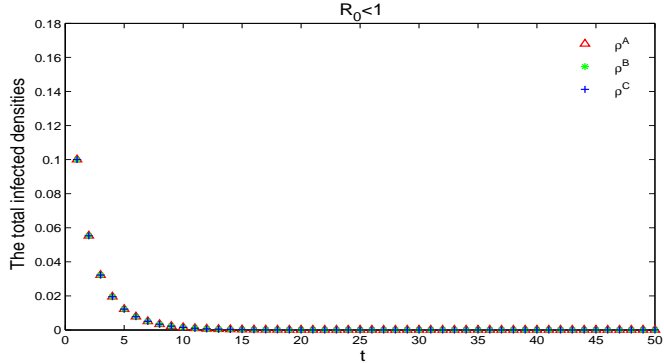}
\caption{The time evolution of the total infected densities. All the infection rates are fixed at $0.15$, and $R_0=0.67<1$.}\label{e003}
\end{center}
\end{figure}

\begin{figure}[]
\begin{center}
\includegraphics[width=10cm,height=5cm]{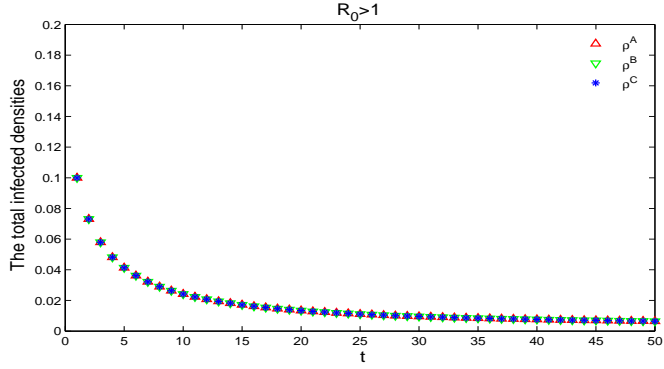}
\caption{The time evolution of the total infected densities. All the infection rates are fixed at $0.25$, and $R_0=1.12>1$.}\label{e004}
\end{center}
\end{figure}

In the simulations, we set $N^A=N^B=N^C=1000,\ k_0=1,\ \gamma=2.8$, and $\mu_1=\mu_2=\mu_3=1$. First, we testify the availability of the basic reproduction $R_0$ obtained by matrix (\ref{e05}). In Fig.~(\ref{e003}), all the infection rates are fixed at $0.15$, and $R_0=0.67<1$, we can see the disease in all subnetworks disappear in the end. In Fig.~(\ref{e004}), all the infection rates are fixed at $0.25$, and $R_0=1.12>1$, the disease spreads and become endemic. In Fig.~(\ref{e003}) and (\ref{e004}), all the contact patterns are scale-free. One can see that the values of $R_0$ are consistent with theoretical result very well.

\begin{figure}[]
\begin{center}
\includegraphics[width=10cm,height=5cm]{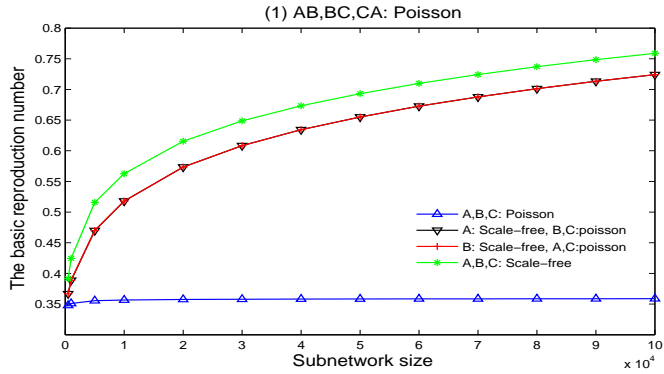}
\caption{The relationship between the basic reproduction number and the network size for different network structures. The networks of AB, BC and CA are poisson.}\label{e005}
\end{center}
\end{figure}

\begin{figure}[]
\begin{center}
\includegraphics[width=10cm,height=5cm]{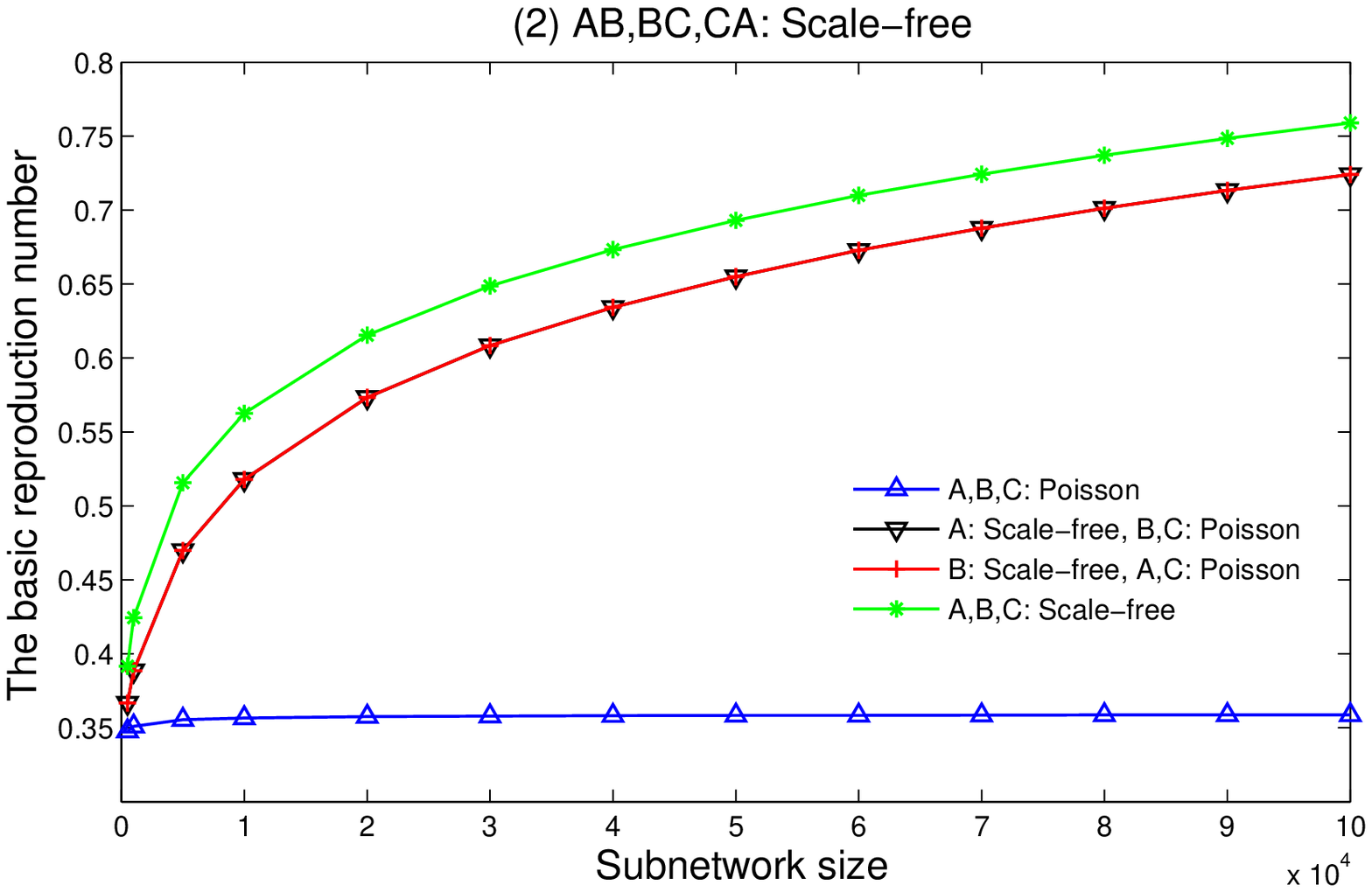}
\caption{The relationship between the basic reproduction number and the network size for different network structures. The networks of AB, BC and CA are scale-free.}\label{e006}
\end{center}
\end{figure}

\begin{figure}[]
\begin{center}
\includegraphics[width=10cm,height=5cm]{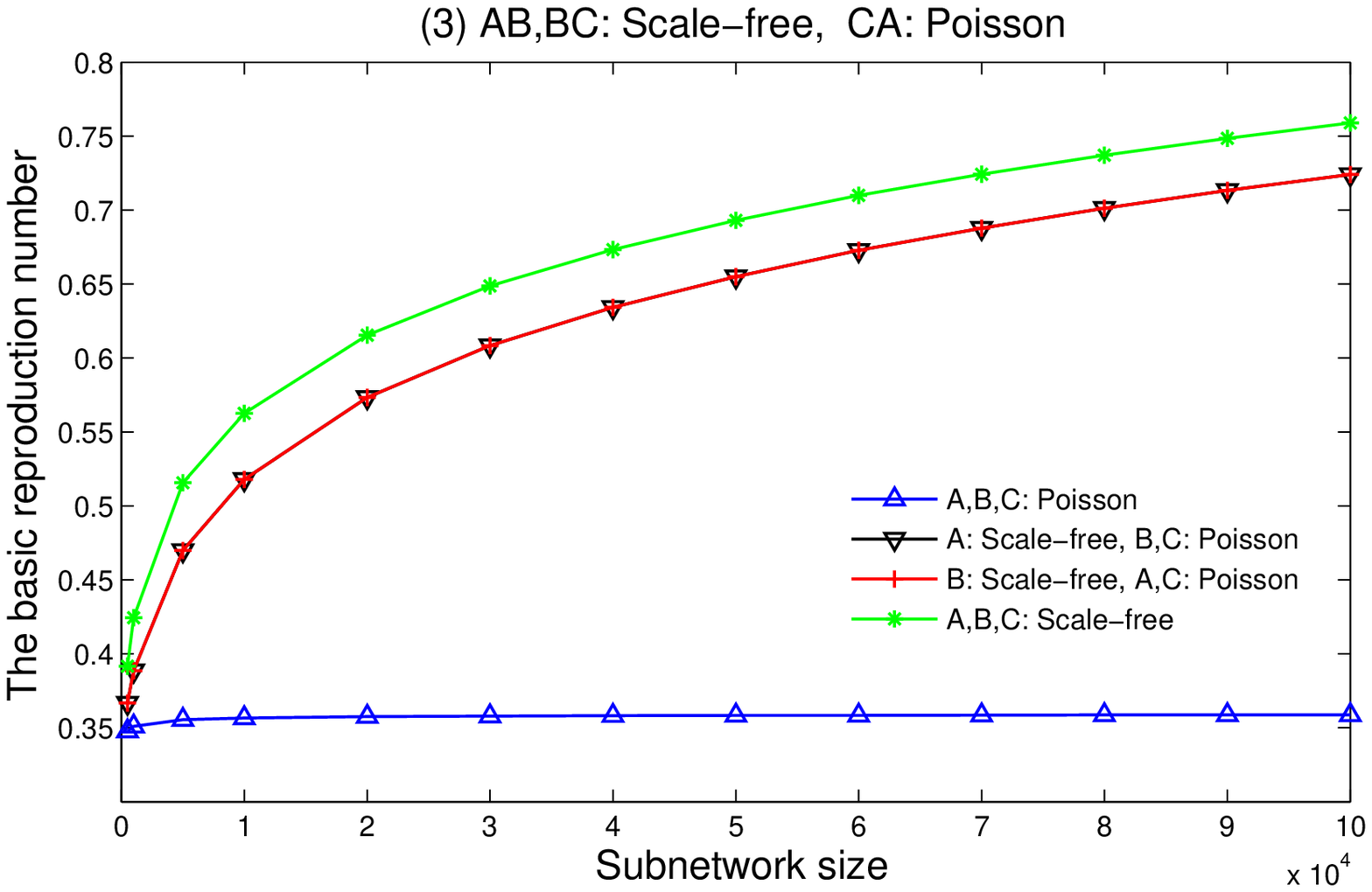}
\caption{The relationship between the basic reproduction number and the network size for different network structures. The networks of AB, BC are scale-free, and the network of CA is poisson.}\label{e007}
\end{center}
\end{figure}

\begin{figure}[]
\begin{center}
\includegraphics[width=10cm,height=5cm]{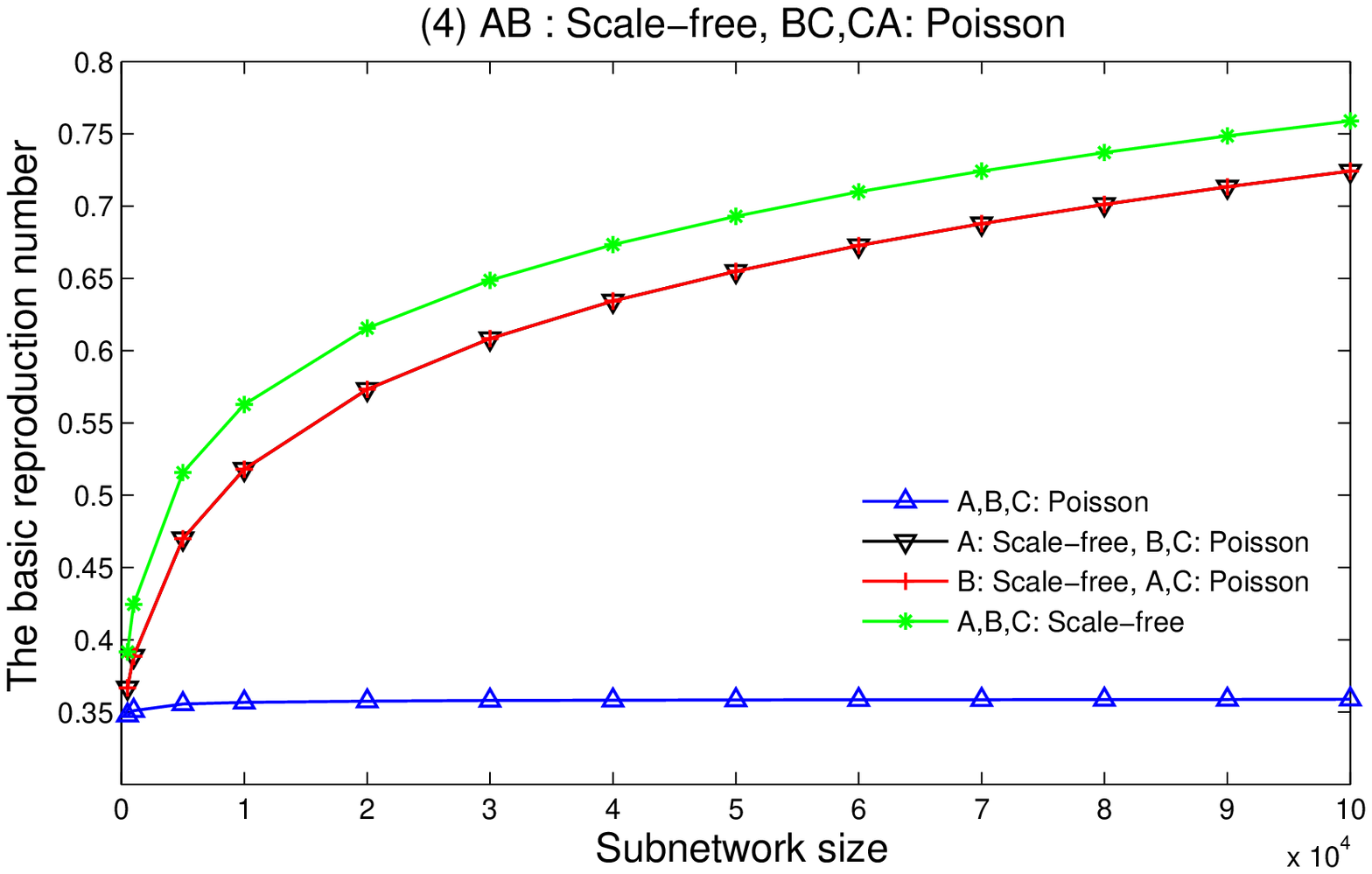}
\caption{The relationship between the basic reproduction number and the network size for different network structures. The network of AB is scale-free, and the networks of BC and CA are poisson.}\label{e008}
\end{center}
\end{figure}

According to Fig.~(\ref{e005})-(\ref{e008}), we analysis the relationship between the basic reproduction number and the network size for different network structures. Hence, all the infection rates are $0.1$, the other parameters be same as the above. All the contact patterns admit the same average degrees. Form the figures, we can observe that a scale-free subnetwork can make $R_0$ increase rapidly with the growth of network size, while a random subnetwork almost do not effect on $R_0$. Moreover, subnetworks $A$, $B$ or $C$ have almost the same impact on $R_0$. Especially, the different network structures of the cross contact AB, BC and CA patterns hardly have any remarkable impact on $R_0$.

\begin{figure}[]
\begin{center}
\includegraphics[width=10cm,height=5cm]{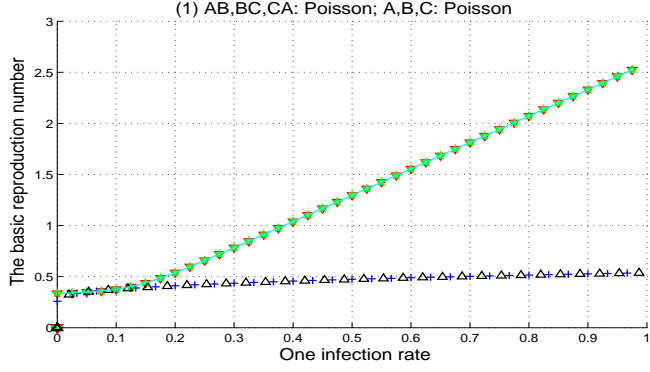}
\caption{The relationship between the basic reproduction number and the infection rates for different network structures. The networks of A, B, C, AB, BC and CA are poisson.}\label{e009}
\end{center}
\end{figure}

\begin{figure}[]
\begin{center}
\includegraphics[width=10cm,height=5cm]{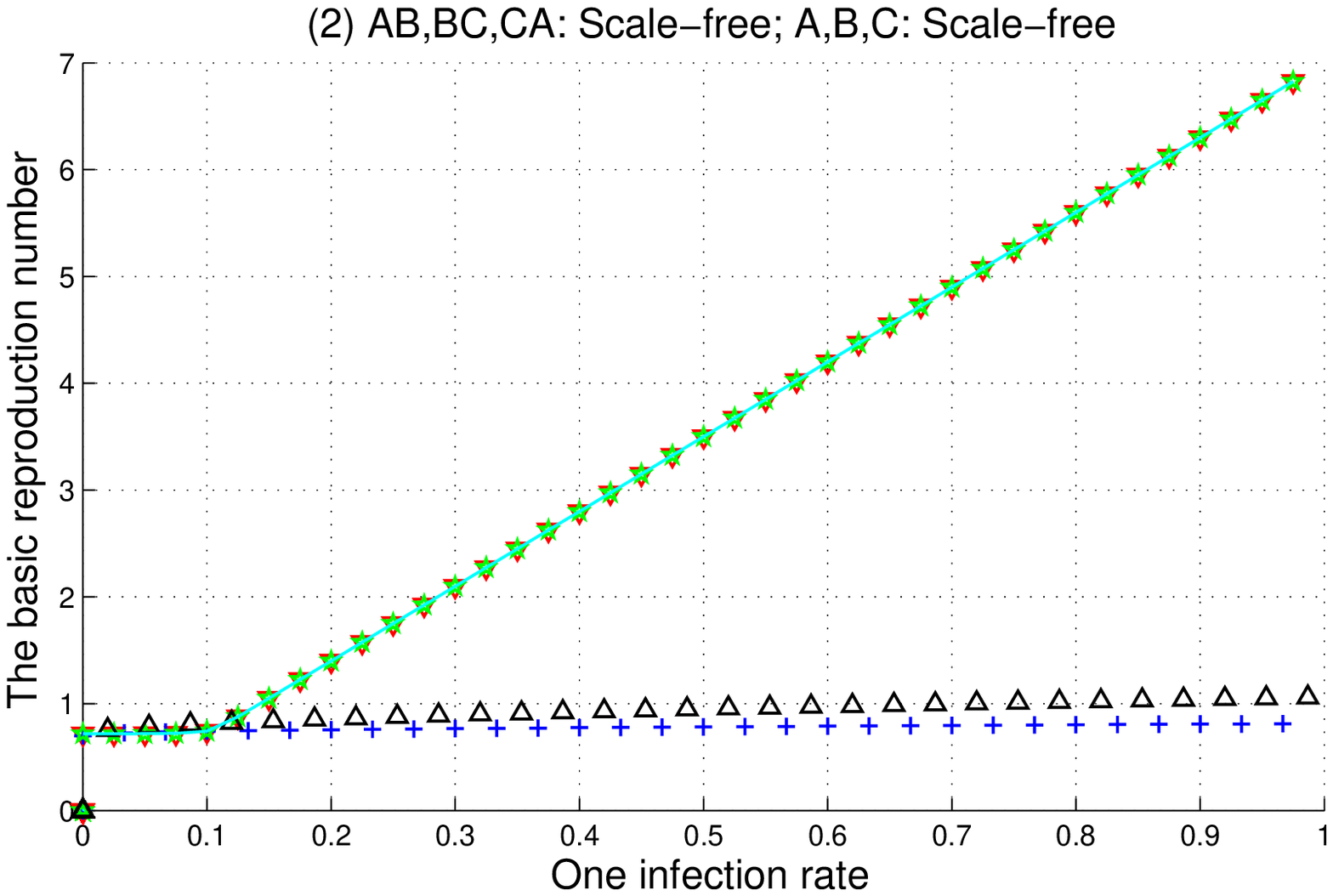}
\caption{The relationship between the basic reproduction number and the infection rates for different network structures. The networks of A, B, C, AB, BC and CA are scale-free.}\label{e0010}
\end{center}
\end{figure}

\begin{figure}[]
\begin{center}
\includegraphics[width=10cm,height=5cm]{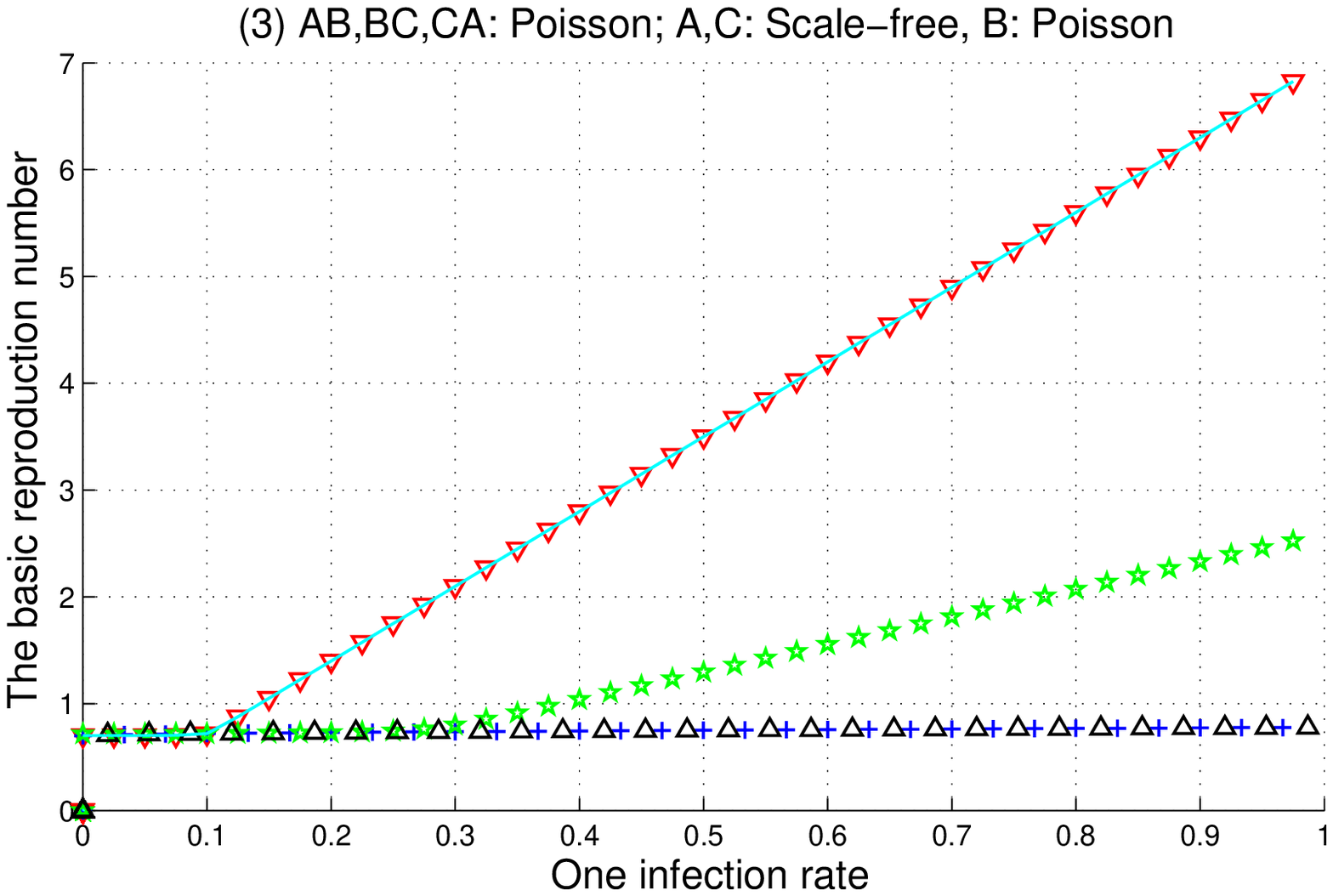}
\caption{The relationship between the basic reproduction number and the infection rates for different network structures. The networks of B, AB, BC and CA are poisson, the networks of A and C are scale-free.}\label{e0011}
\end{center}
\end{figure}

\begin{figure}[]
\begin{center}
\includegraphics[width=10cm,height=5cm]{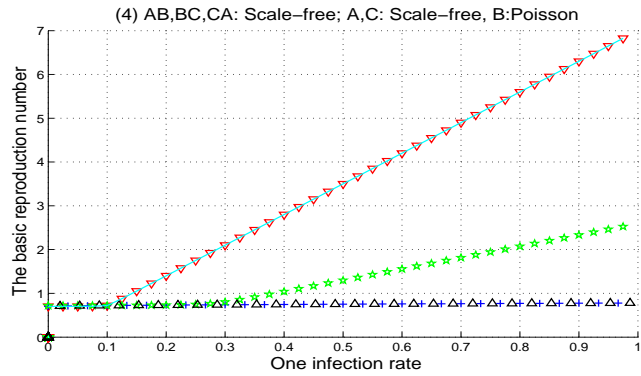}
\caption{The relationship between the basic reproduction number and the infection rates for different network structures. The networks of A, C, AB, BC and CA are scale-free, the network of B is poisson.}\label{e0012}
\end{center}
\end{figure}

Then, we continue to address how the infection rates affect $R_0$ on different network structures. In Fig.~(\ref{e009})-(\ref{e0012}), we set $N^A=N^B=N^C=10000$, when one of the infection rates changes, the others are fixed at $0.1$. The other parameters be same as the above. The label on the x-axis in $\lambda_{11},\ \lambda_{12},\ \lambda_{22},\ \lambda_{31}$ and $\lambda_{33}$ with the corresponding the red $\triangledown$, blue $+$, green $*$, black $\vartriangle$ and cyan $-$, respectively. All the contact patterns admit the same average degrees. One can see that under the same contact pattern, the internal infection rates $\lambda_{11}$, $\lambda_{22}$ and $\lambda_{33}$ have the same influence on $R_0$, as well as the cross-infection rates $\lambda_{12}$, $\lambda_{23}$ and $\lambda_{31}$, and the internal infection rates have greater influence on $R_0$ than the cross-infection rates. So£¬the internal infection can effect the epidemic more than the cross-infection. If one of the inner contact patterns is scale-free, then $R_0$ increases quickly with the growth of inner infection rate.

\begin{figure}[]
\begin{center}
\includegraphics[width=10cm,height=5cm]{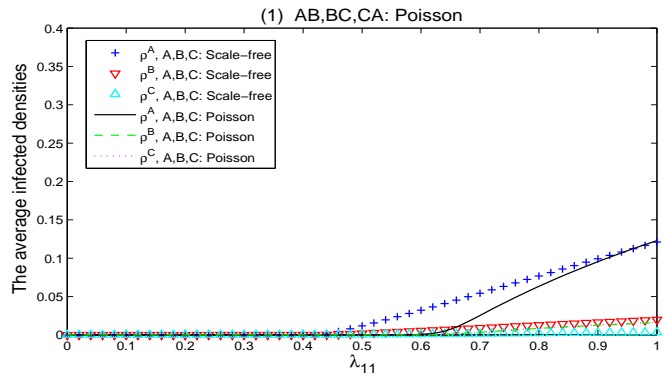}
\caption{The average infected densities depend on the infection rates and the contact patterns. When one of the infection rates changes, the others are fixed at $0.1$.}\label{e0013}
\end{center}
\end{figure}

\begin{figure}[]
\begin{center}
\includegraphics[width=10cm,height=5cm]{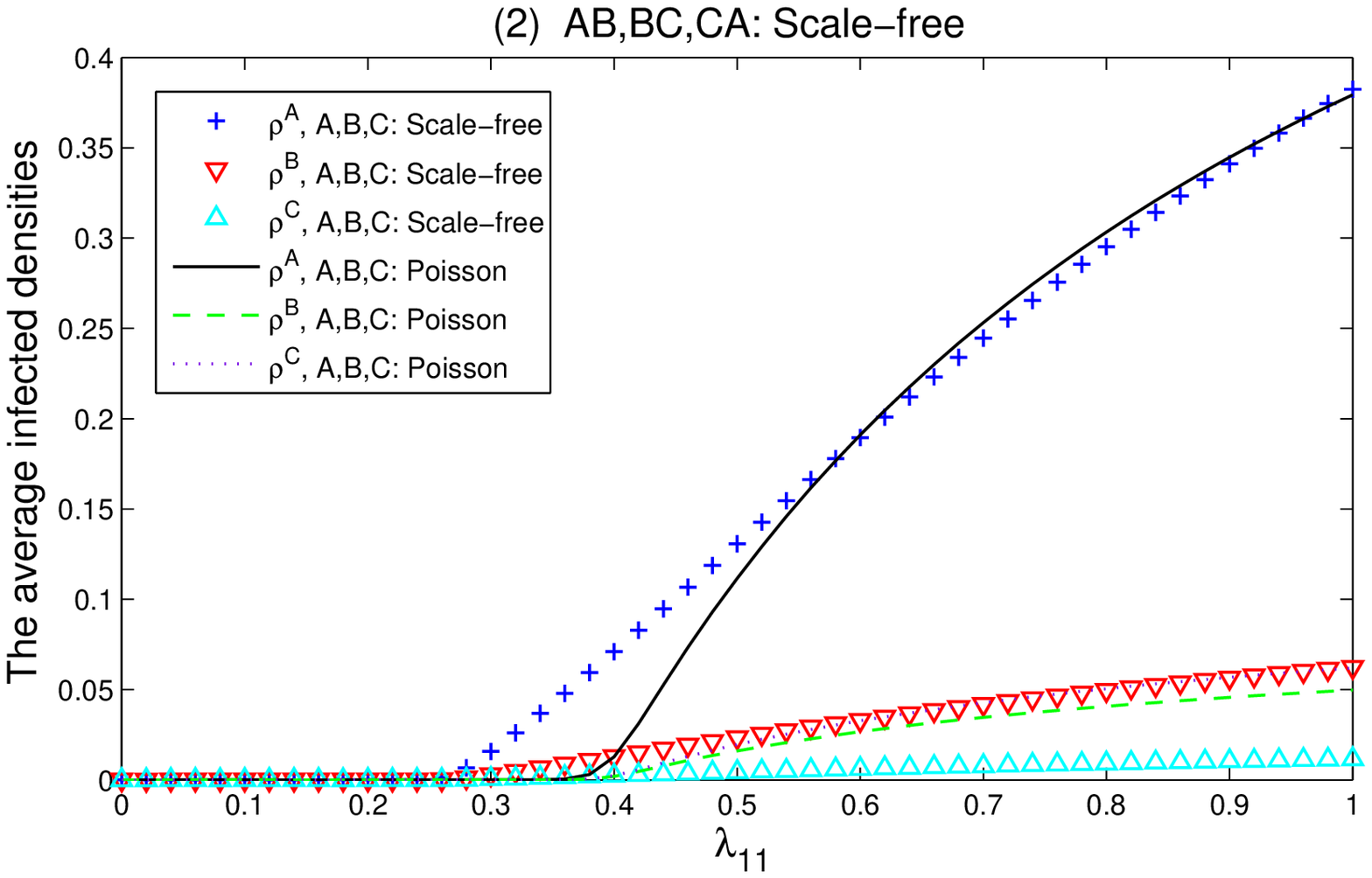}
\caption{The average infected densities depend on the infection rates and the contact patterns. When one of the infection rates changes, the others are fixed at $0.1$.}\label{e0014}
\end{center}
\end{figure}

\begin{figure}[]
\begin{center}
\includegraphics[width=10cm,height=5cm]{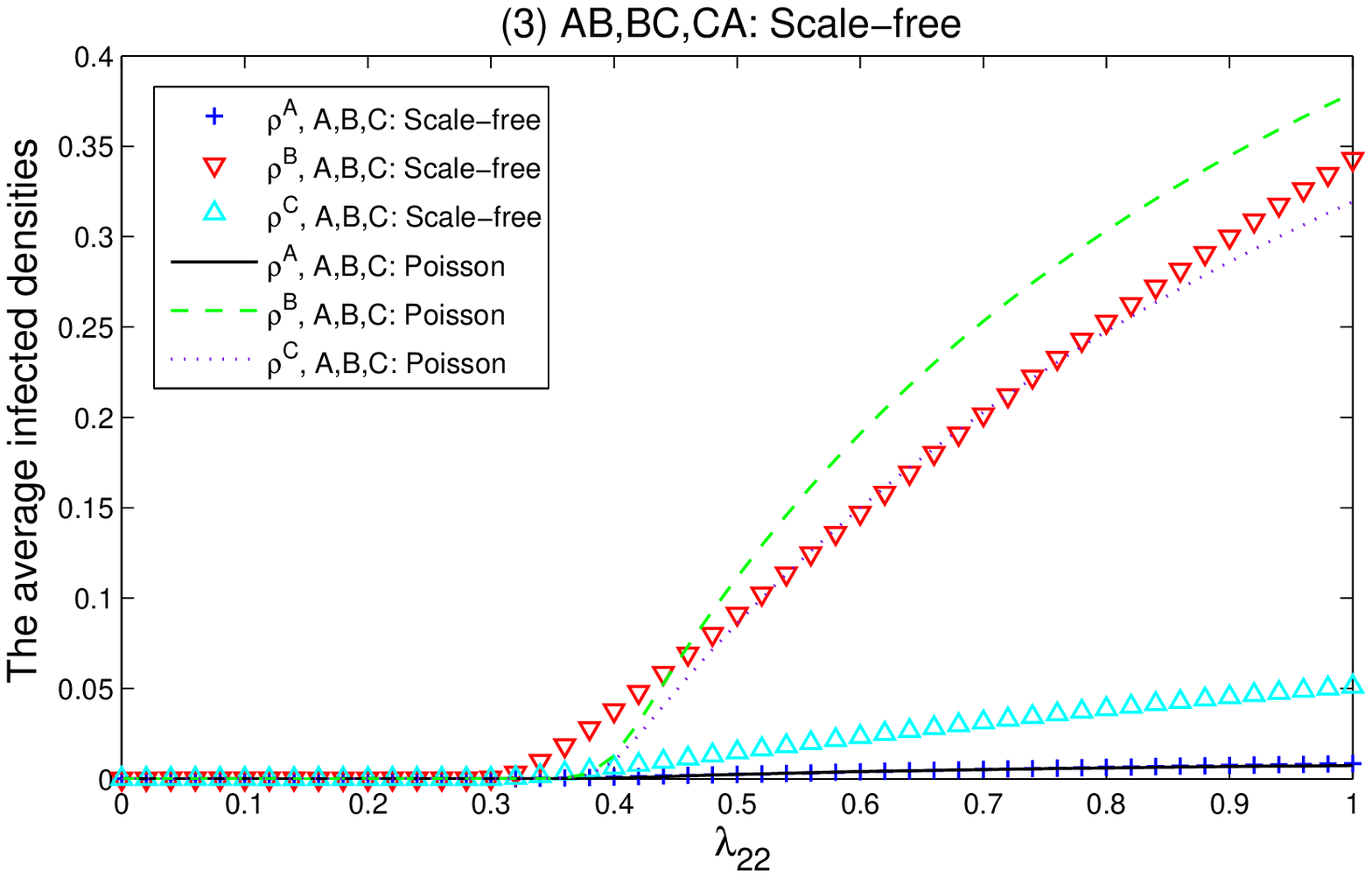}
\caption{The average infected densities depend on the infection rates and the contact patterns. When one of the infection rates changes, the others are fixed at $0.1$.}\label{e0015}
\end{center}
\end{figure}

\begin{figure}[]
\begin{center}
\includegraphics[width=10cm,height=5cm]{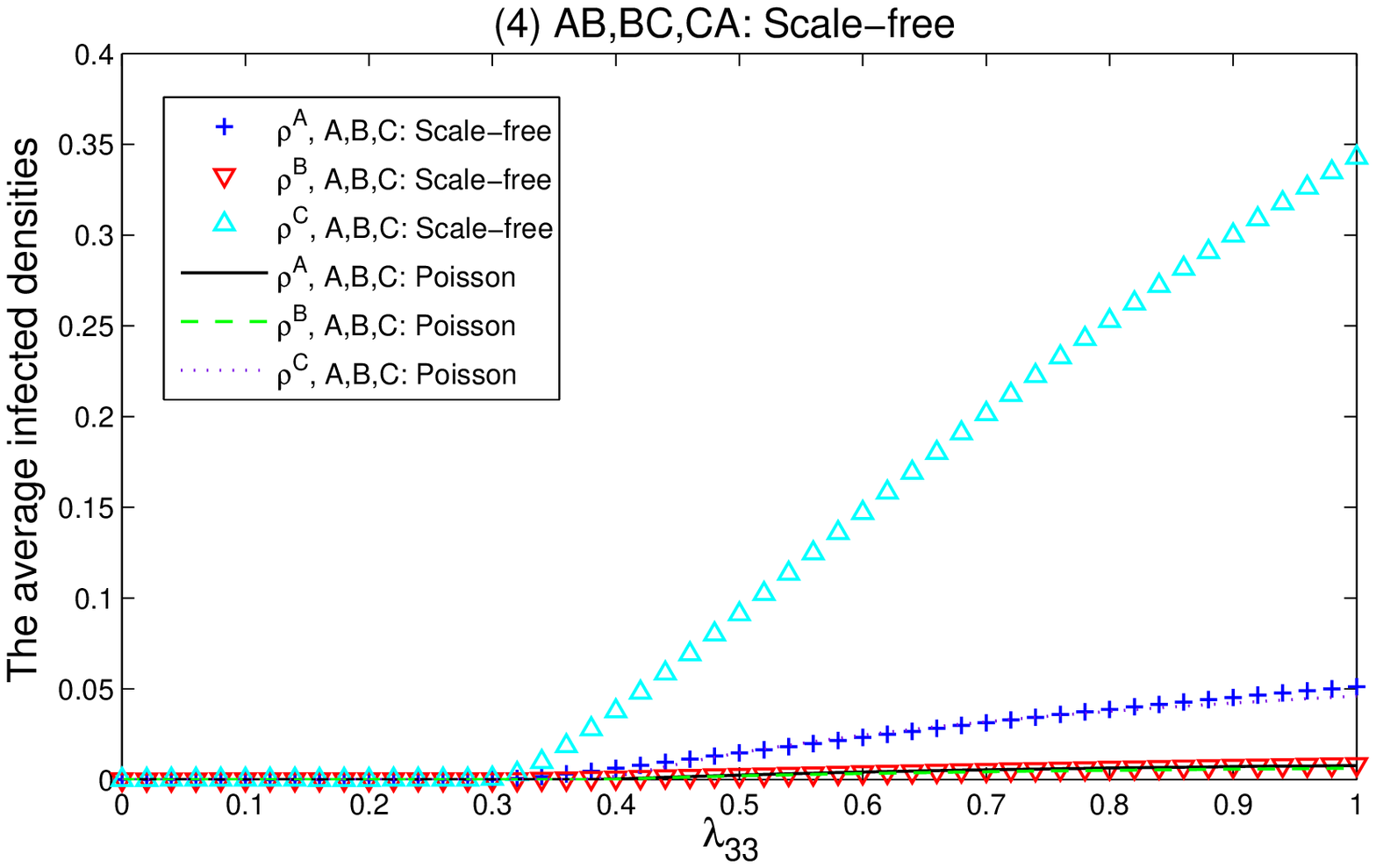}
\caption{The average infected densities depend on the infection rates and the contact patterns. When one of the infection rates changes, the others are fixed at $0.1$.}\label{e0016}
\end{center}
\end{figure}

Finally, we also set $N^A=N^B=N^C=1000,\ k_0=1,\ \gamma=2.8$, and $\mu_1=\mu_2=\mu_3=1$ in Fig.~(\ref{e0013})-(\ref{e0016}), when one of the infection rates changes, the others are fixed at $0.1$. One can observe that the scale-free contact pattern can make the bigger average infected densities and the smaller epidemic thresholds for the infected rates than the random contact pattern. With the inner infection rate $\lambda_{11}$ increase, the average infected density of subnetwork $A$ increases rapidly, followed by the infected density of subnetwork $B$, but almost no reaction with subnetwork $C$. It is similar to the increase of the infected rates $\lambda_{22}$ and $\lambda_{33}$.

\subsection{One-way circular-coupled network with infective media}

From the system $(\ref{e06})$, we know that the infection rates and the network structures have a impact on the dynamical behavior of the infectious disease. As we all know, contact pattern of human network is heterogeneous, while animal's contact pattern is homogeneous. We suppose there are two human subnetworks and an animal subnetwork in the network. So, we set a BA scale-free and ER random network as the topological structure of the human network and the animal network, respectively. Let subnetworks $A$, $B$ and $C$ have the same nodes $1000$, the subnetworks $A$ and $B$ are BA scale-free networks and the subnetwork $C$ is ER random network, all the same average degrees $\langle{k}\rangle=\langle{l}\rangle=\langle{m}\rangle=4$.

\begin{figure}[]
\begin{center}
\includegraphics[width=10cm,height=5cm]{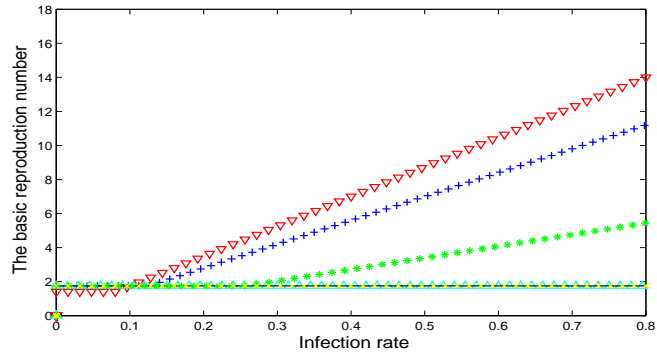}
\caption{The influence of various infection rates on the basic reproduction number $R_0$. When one of the infection rates changes, the others are fixed to be $0.1$.}\label{e0017}
\end{center}
\end{figure}

From Fig.~(\ref{e0017}), one can see that the influence of various infection rates on the basic reproduction number $R_0$. The line $+$, $\triangle$, $\nabla$, $-$, $*$ and $\times$ denote the change of $\lambda_1$, $r_{32}$, $\lambda_2$, $r_{11}$, $\lambda_3$ and $r_{12}$. It is found that $\lambda_2$ contributes most to the basic reproduction number, $\lambda_1$ counts second, $\lambda_3$ counts third, and $r_{11}$, $r_{12}$ and $r_{32}$ have hardly impact on it. The disease prevails in a heterogeneous network has a greater impact on $R_0$ than the disease from a homogeneous network.

\begin{figure}[]
\begin{center}
\includegraphics[width=10cm,height=5cm]{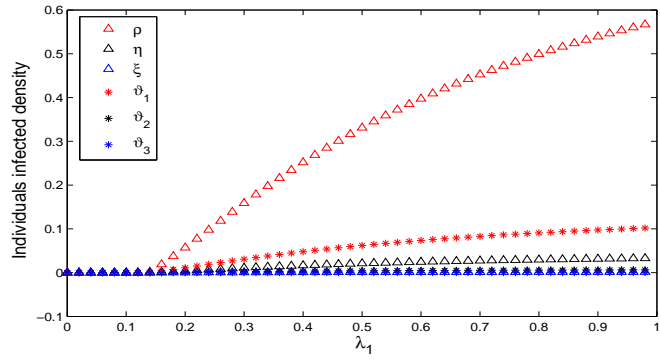}
\caption{The relationship between individuals infected density and the infection rate $\lambda_1$, where $\lambda_2=\lambda_3=0.2$.}\label{e0018}
\end{center}
\end{figure}

\begin{figure}[]
\begin{center}
\includegraphics[width=10cm,height=5cm]{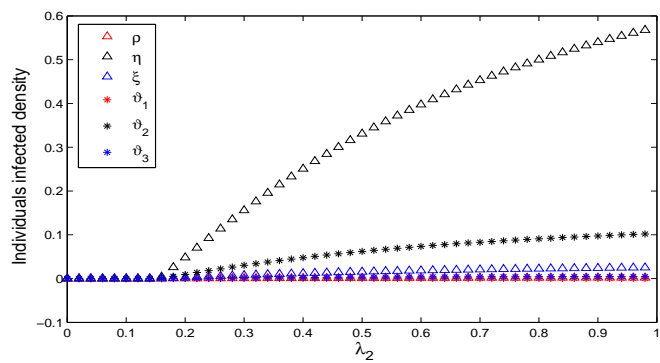}
\caption{The relationship between individuals infected density and the infection rate $\lambda_2$, where $\lambda_1=\lambda_3=0.2$.}\label{e0019}
\end{center}
\end{figure}

Next, Fig.~(\ref{e0018})-(\ref{e0019}) give the influence of infection rates on the stable infected density. We set $r_{11}=r_{12}=r_{21}=r_{22}=r_{31}=r_{32}=0.1$. The inner infection rates $\lambda_2=\lambda_3=0.2$ in Fig.~(\ref{e0018}) and $\lambda_1=\lambda_3=0.2$ in Fig.~(\ref{e0019}), respectively. From Fig.~(\ref{e0018}), it is observed that the threshold admits the same value for these populations, with the infection rate $\lambda_1$ increase, the infected density $\rho$ of subnetwork $A$ increases rapidly, more slowly in infected density $\eta$ and $\vartheta_1$, but almost no reaction in $\xi$, $\vartheta_2$ and $\vartheta_3$. Since the direction of infectious diseases transmission between population networks and vectors is one-way, subnetwork $A$ almost has no influence on subnetwork $C$, vectors $b$ and $c$. It is similar to Fig.~(\ref{e0019}), along with the growth of infection rate $\lambda_2$, the infected density $\eta$ of subnetwork $B$ increases rapidly, more slowly in infected density $\xi$ and $\vartheta_2$, but almost no reaction in $\rho$, $\vartheta_1$ and $\vartheta_3$.

\begin{figure}[]
\begin{center}
\includegraphics[width=10cm,height=5cm]{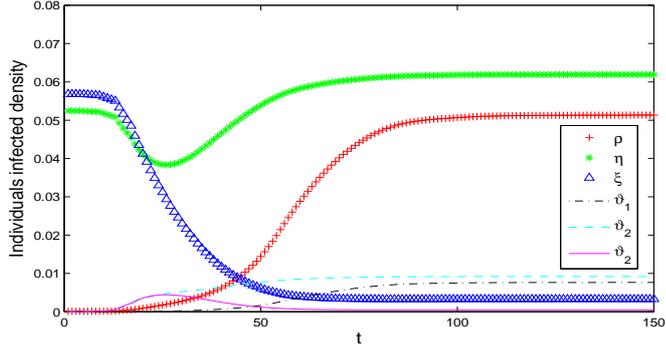}
\caption{The time behavior of the infected density, where $r_{11}=r_{12}=r_{21}=r_{22}=r_{31}=r_{32}=0.15$ and $R_0=1.5$.}\label{e0020}
\end{center}
\end{figure}

\begin{figure}[]
\begin{center}
\includegraphics[width=10cm,height=5cm]{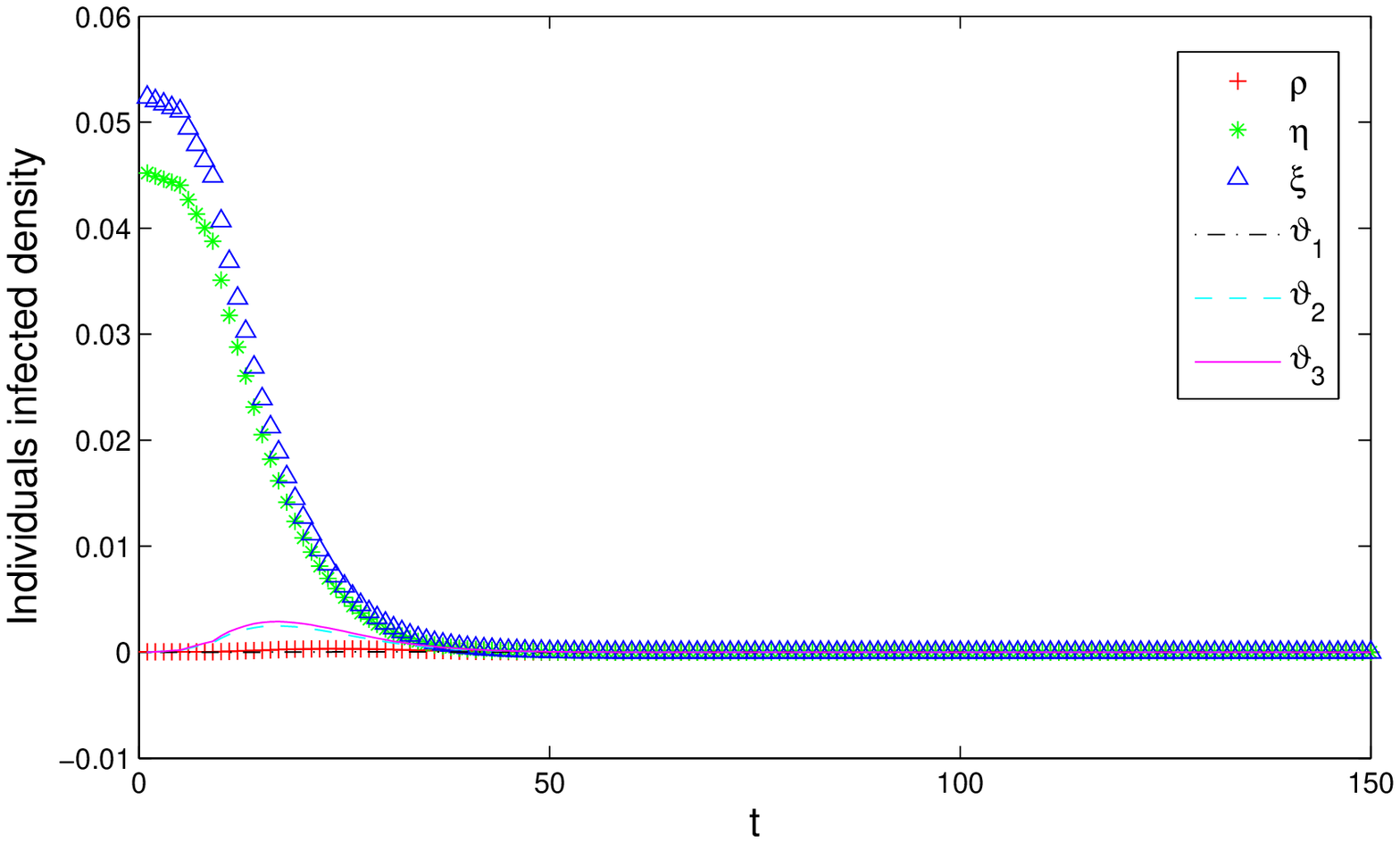}
\caption{The time behavior of the infected density, where $r_{11}=r_{12}=r_{21}=r_{22}=r_{31}=r_{32}=0.15$ and $R_0=0.15$.}\label{e0021}
\end{center}
\end{figure}

\begin{figure}[]
\begin{center}
\includegraphics[width=10cm,height=5cm]{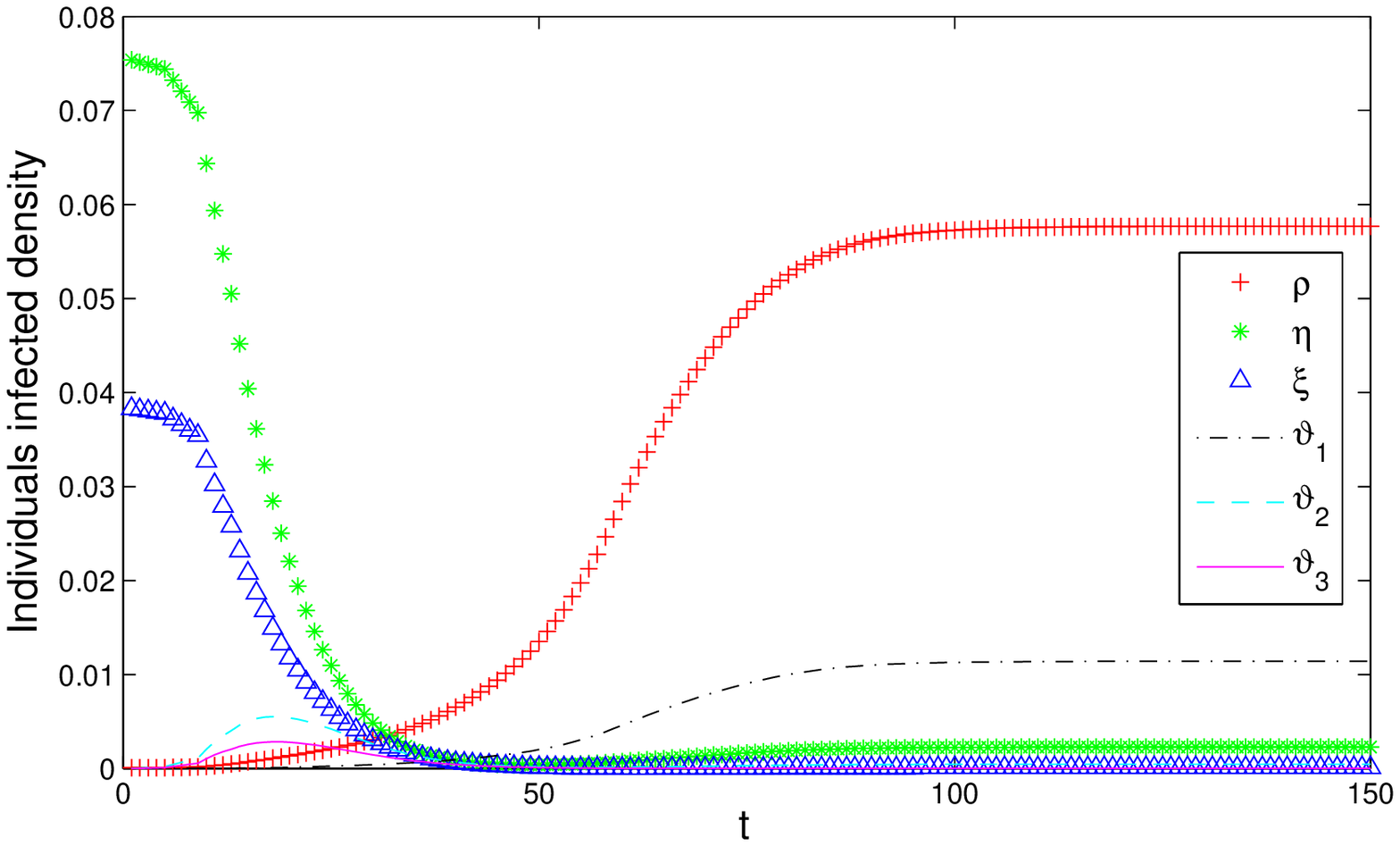}
\caption{The time behavior of the infected density, where $r_{11}=r_{12}=r_{21}=r_{22}=r_{31}=r_{32}=0.2$. When $\lambda_1=0.2$ and $\lambda_2=\lambda_3=0$, then $R_0=2$.}\label{e0022}
\end{center}
\end{figure}

\begin{figure}[]
\begin{center}
\includegraphics[width=10cm,height=5cm]{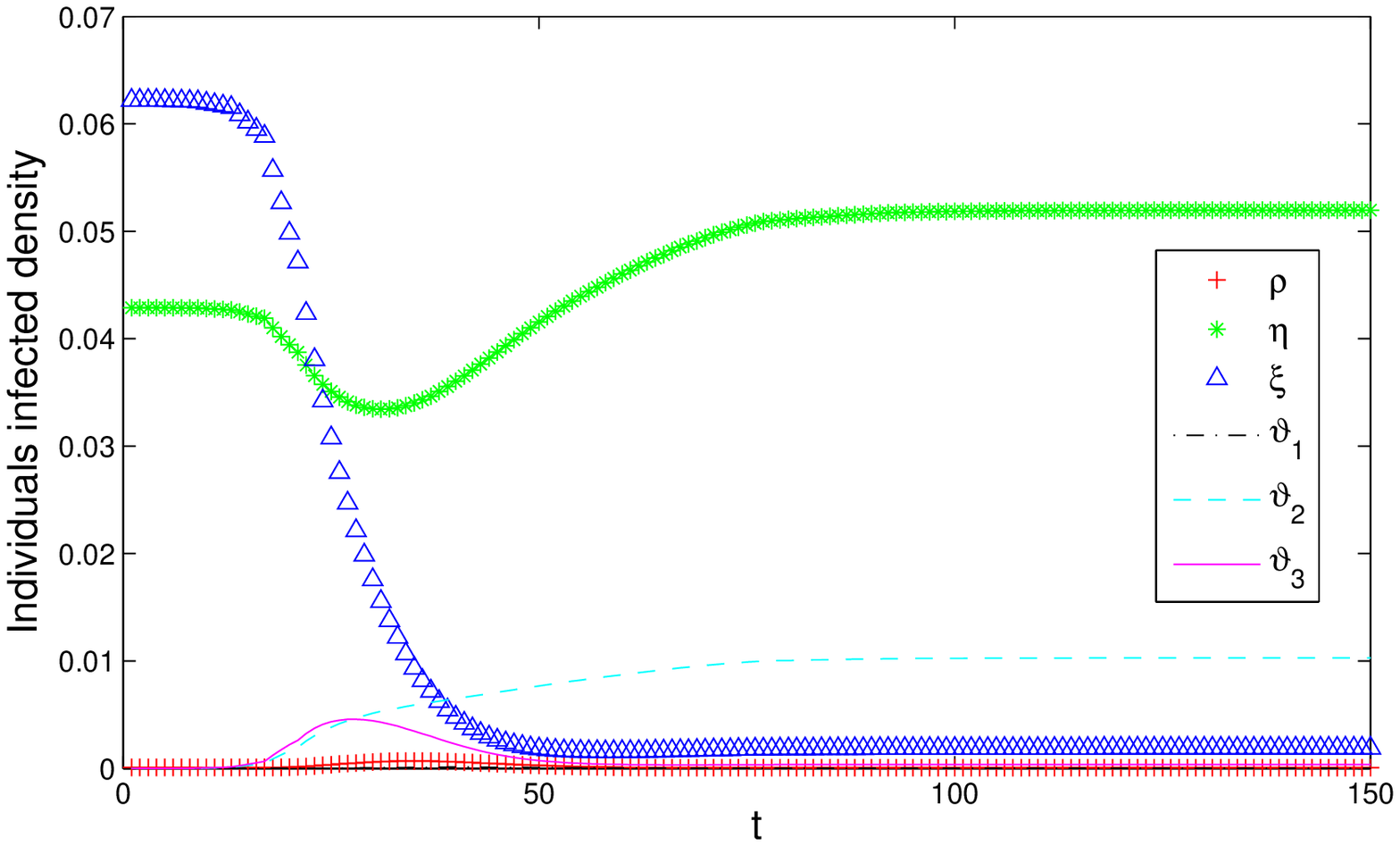}
\caption{The time behavior of the infected density, where $r_{11}=r_{12}=r_{21}=r_{22}=r_{31}=r_{32}=0.2$. When $\lambda_2=0.2$ and $\lambda_1=\lambda_3=0$, then $R_0=2$.}\label{e0023}
\end{center}
\end{figure}

\begin{figure}[]
\begin{center}
\includegraphics[width=10cm,height=5cm]{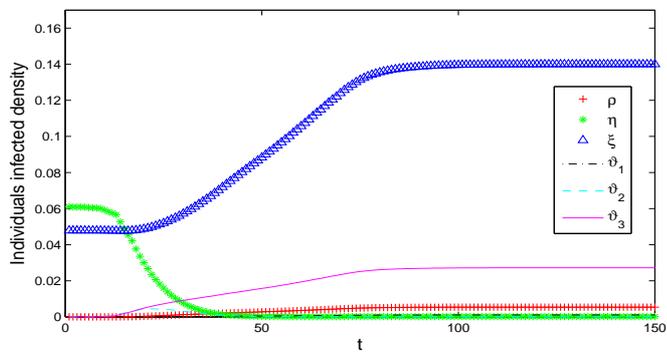}
\caption{The time behavior of the infected density, where $r_{11}=r_{12}=r_{21}=r_{22}=r_{31}=r_{32}=0.2$. When $\lambda_3=0.25$ and $\lambda_1=\lambda_2=0$, then $R_0=1.25$.}\label{e0024}
\end{center}
\end{figure}

The time behaviors of the infectious disease propagation are described in Fig.~(\ref{e0020})-(\ref{e0024}). In Fig.~(\ref{e0020}), we set $\lambda_1=\lambda_2=\lambda_3=0.15$, which imply all the populations are infective, after a process of slow propagation, the infected density in subnetwork $A$ grows rapidly, the infection scale in subnetwork $B$ begins to drop, then increase and reach the stable state, but the incidence in subnetwork $C$ gradually becomes a weaker stable state, which is attribute to inner infection of subnetwork $A$ and $B$ as the dominant role. Infected densities of all the vectors reach the lower stable state. In Fig.~(\ref{e0021}), we consider the disease just spread by the vectors, so we set $\lambda_1=\lambda_2=\lambda_3=0$, then the basic reproduction number $R_0<1$, which imply the disease would quickly disappear. One can see the disease in subnetwork $B$ and $C$ are decrease rapidly, and even no incidence in subnetwork $A$. We set $\lambda_1=0.2$, $\lambda_2=\lambda_3=0$ in Fig.~(\ref{e0022}), which means the inner infection only in subnetwork $A$. The disease in $A$ increase rapidly and become stable state, the incidences in $B$ and $C$ both reach a low stable state, as $\lambda_1$ increases, the disease gradually across the vector $a$ and then infect the individuals in subnetwork $B$. Hence, we set $\lambda_2=0.2$, $\lambda_1=\lambda_3=0$ in Fig.~(\ref{e0023}) and $\lambda_3=0.25$, $\lambda_1=\lambda_2=0$ in Fig.~(\ref{e0024}), the conclusion also like above. In this case, the infections among the three subnetworks all play a important role in the propagation process, due to the particularity of transmission, the disease in one subnetwork has the bigger impact on the next vector, followed by the next subnetwork and almost no impact on another subnetwork.

\begin{figure}[]
\begin{center}
\includegraphics[width=10cm,height=5cm]{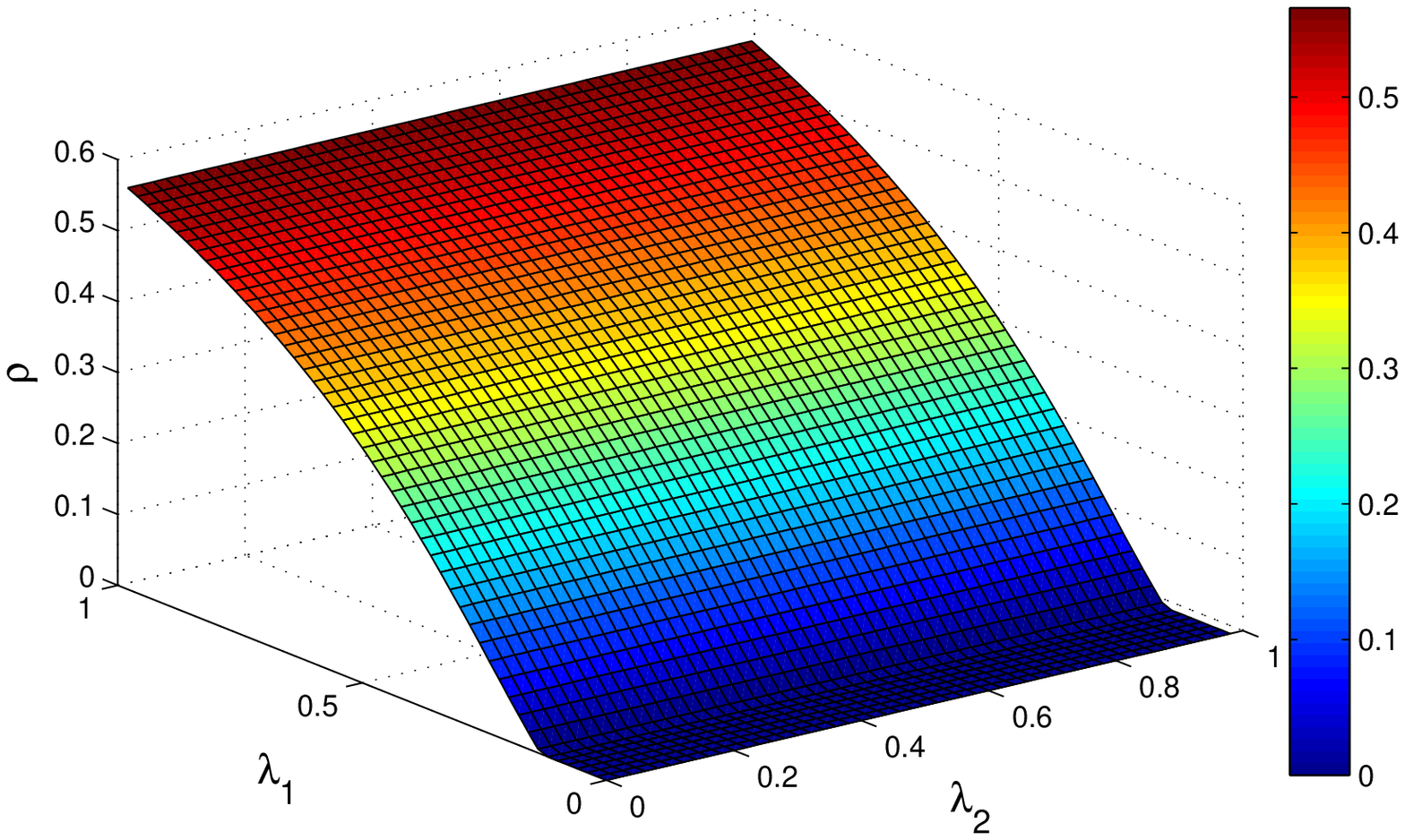}
\caption{The stable infected density of network versus the transmission coefficients. When the parameters change, others are fixed on 0.05.}\label{e0025}
\end{center}
\end{figure}

\begin{figure}[]
\begin{center}
\includegraphics[width=10cm,height=5cm]{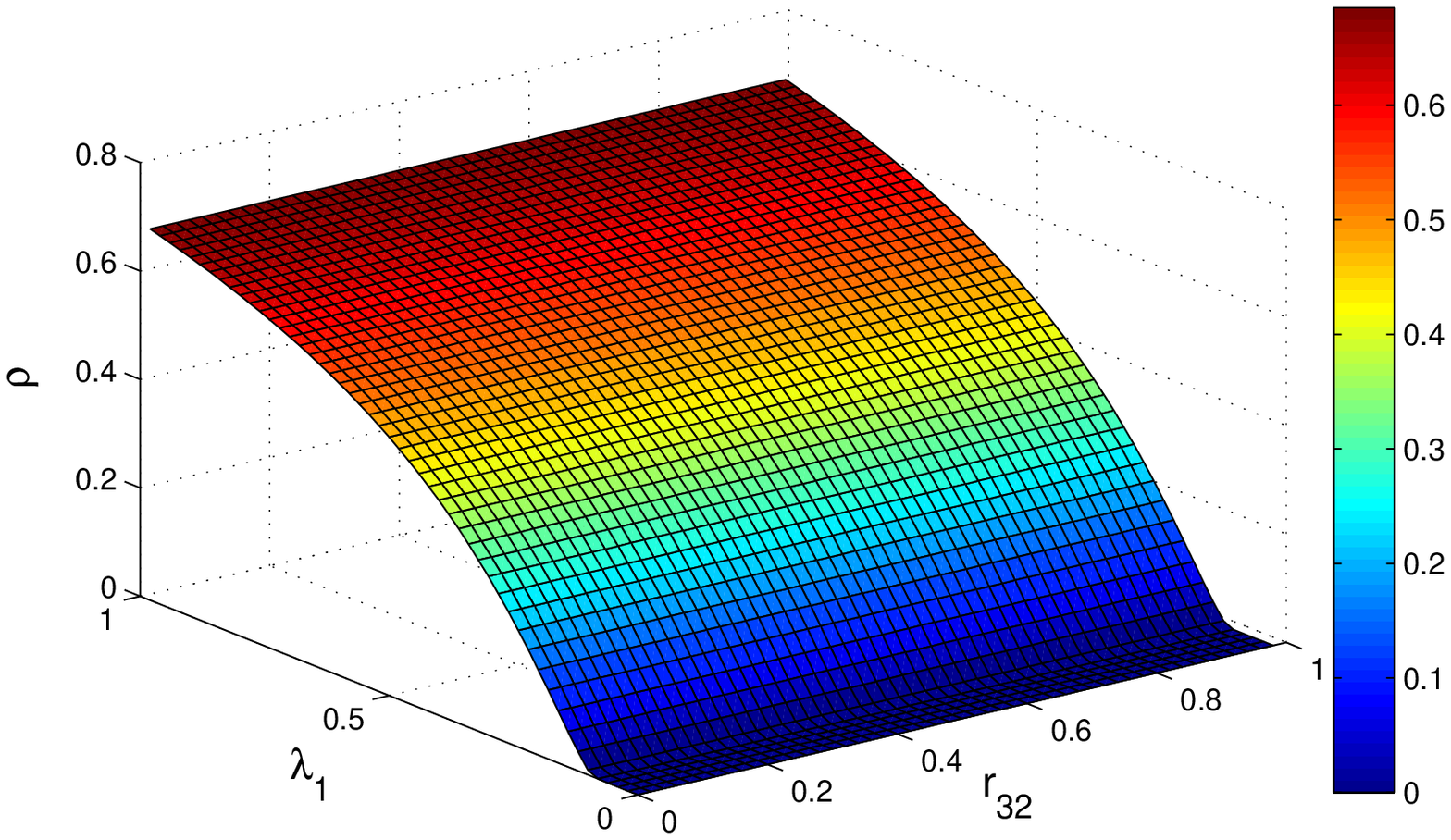}
\caption{The stable infected density of network versus the transmission coefficients. When the parameters change, others are fixed on 0.05}\label{e0026}
\end{center}
\end{figure}

\begin{figure}[]
\begin{center}
\includegraphics[width=10cm,height=5cm]{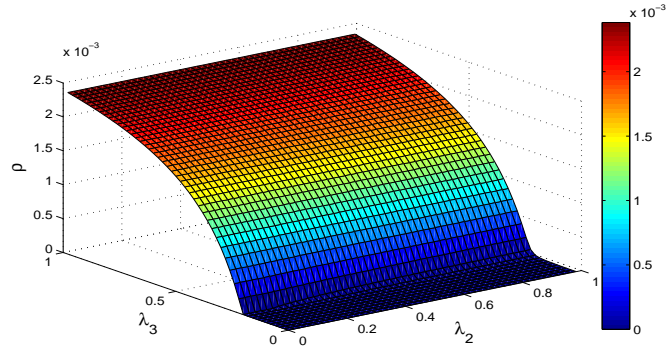}
\caption{The stable infected density of network versus the transmission coefficients. When the parameters change, others are fixed on 0.05}\label{e0027}
\end{center}
\end{figure}

Finally, Fig.~(\ref{e0025})-(\ref{e0027}) show the infected density in subnetwork $A$ as a function of the transmission coefficients. It is found that if the inner infection rate $\lambda_1$ ($\lambda_2$) in subnetwork $A$ ($B$) is weak, then the infection from vector $c$ almost no effect on subnetwork $A$. When the infection rate $\lambda_2$ is fixed, the infected density $\rho$ increases with the growth of $\lambda_1$. If the inner infection rate $\lambda_1$ is small, the cross infection $r_{32}$ has little effect on the infected density. When the infection rates $\lambda_2$ and $\lambda_3$ are very strong, but the endemic stable density of network $A$ is very small. So, $\lambda_2$ and $\lambda_3$ generate less incidence on the disease transmission. The simulation results are independent of the initial conditions, in accordance with the theoretical results.

\section{Conclusions and discussions}

In this paper, to study the epidemic spreading on three different populations with different contact patterns and infection rates, and the zoonotic infections spread on interconnected networks, we established and investigated two one-way three-layer circular interdependent networks. The first epidemic model contains three different populations, the disease spread is one-way, then becomes a circular-coupled network. There are three different infective medias are embedded in the second model. By the heterogeneous mean-field approach method, we estimate the basic reproduction number $R_0$ of these two models by Perron-Frobenius theorem, then give the proof of the global stability of the disease-free equilibrium and endemic equilibrium of the first model. In the second model, we proved that when $R_0<1$, the disease-free equilibrium is globally asymptotically state, if $R_0>1$, the unique endemic equilibrium is globally attractive.

Through mathematical analysis and numerical simulations, we can obtain some significant results. In the first model, all the inner and across interactions can contribute to the epidemic spreading, the inner contact patterns have a bigger impact on the basic reproduction number $R_0$ than the cross contact patterns. Under the same contact pattern, the internal infection rates have greater influence on $R_0$ than the cross-infection rates. Moreover, three subnetworks have almost the same impact on $R_0$, and the different cross contact patterns hardly have any remarkable impact on $R_0$. As a general situation, we set two human subnetworks and an animal subnetwork in the second model. The disease prevails in a heterogeneous network has a greater impact on $R_0$ than the disease from a homogeneous network. With the increase of the infection rate $\lambda_1$, subnetwork $A$ has the bigger impact on vector $a$, the smaller effect on subnetwork $B$, almost no influence on network $C$, vectors $b$ and $c$. Other similar to this. Due to the particularity of transmission, from the simulation results we obtain that the disease in one subnetwork has the bigger impact on the next vector, followed by the next subnetwork and almost no impact on another subnetwork. One can see that if a disease has weak inner infection in subnetwork, then the infection from vectors barely impact the subnetwork.

Hence, this work gives two special one-way circular-coupled epidemic models for better understanding and some new viewpoints for the research of epidemic spread in real life. Our results will provide certain theoretical support for real applications in human infectious disease control.

\addcontentsline{toc}{chapter}{Acknowledgment}
\subsection*{Acknowledgment}

This work was jointly supported by the NSFC grants 11572181 and 11331009, and partly done while visiting the Department of Computer and Information Science, Univ. of Macau. We'd like to thank Prof. Stephen Gourley at Univ. of Surrey for his kind help and encouragement in improving the manuscript.

\end{document}